\documentclass[12pt]{article}

\usepackage{feynmp}
\usepackage{slashed}
\usepackage{epsfig}
\usepackage{amsfonts}

\title{The Standard Model Coupled to Quantum Gravitodynamics}

\author{Fermin Aldabe\\ [5mm]
}
\date{ }
\begin{document}
\begin{titlepage}
\maketitle
\thispagestyle{empty}
\begin{abstract}
We show that the renormalizable $SO(4)\times U(1)\times SU(2)\times 
SU(3)$ Yang Mills coupled to matter and the Higgs field fits all the experimentally observed
differential cross sections known in nature. This extended Standard Model reproduces
the experimental gravitational differential cross sections without resorting to the 
graviton field and instead by exchanging $SO(4)$ gauge fields.
By construction, each SO(4) generator in quantum gravitodynamics does not commute with the Dirac gamma matrices. This
produces additional interactions absent to non-Abelian gauge fields in the Standard Model.
The contributions from these new terms yield differential cross sections
consistent with the Newtonian and post Newtonian interactions derived from General Relativity.
Dimensional analysis of the Lagrangian 
shows that all its terms have total dimensionality four or less and therefore that all 
physical quantities in the theory renormalize by finite amounts. 
These properties make QGD the only renormalizable 4-dimensional theory describing gravitational interactions.
\end{abstract}
\end{titlepage}

\section{Introduction}

The structure and 
success of General Relativity have lead us to the believe that the spin-2 graviton field mediates quantum gravitational interactions;
but to date this believe has eluded experimental detection. 
Furthermore, the quantum limit of General Relativity fails to renormalize properly forcing us to further believe that General Relativity
comes from a classical projection of a 10 dimensional string. 
On the other hand, we know that the geometrical object that describes the geodesics in General Relativity is the connection and not the metric
which enters that construction only after arbitrarily imposing the metric compatibility constraint.
Then, we can in practice construct a quantum theory of gravity where the fundamental field is the connection as in\cite{yang} 
or a similar object as we do here.
Nothing but experimental observation constrains these possibilities. Therefore, we 
should not force the graviton upon ourselves and instead focus on matching the proposed model's differential cross sections to
experimental evidence; more so, when we must yet detect the particle mediating gravitational interactions.

Here we present quantum gravitodynamics, QGD, the only renormalizable 4-dimensional theory that reproduces the interactions 
in General Relativity to $\mathcal{O}(\frac{v^2}{c^2})$ and couples to the Standard Model through a covariant derivative 
like the other three interactions in nature.
As we show below, all diagrams in QGD match those of it counterpart in General Relativity to $\mathcal{O}(\frac{v^2}{c^2})$
and the potential that determines a test particle's classical trajectories, are identical for both theories to that same order.
As opposed to General Relativity which does not renormalize, QGD has a metric void of any dynamics, explicitly lacks 
a graviton field or any spin-2 state, and instead uses SO(4) Yang Mills fields 
coupled to spinor singlets to mediate the gravitational interactions. 
Furthermore, QGD reproduces all diagrams 
in General Relativity to $\mathcal{O}(\frac{v^2}{c^2})$ and therefore all experimental evidence  to that same order.
However, as opposed to General Relativity, the terms in the QGD Lagrangian involve products of fields and their derivatives with total dimensionality
of at most four  and therefore renormalization follows\cite{wein2}. 
Thus, QGD is renormalizable, and, like General Relativity, constructed in 4 dimensions. 
Because QGD is defined in terms of Yang Mills fields, it also couples naturally to
the Standard Model to produce a Unified Model with $SO(4)\times U(1)\times SU(2)\times SU(3)$ gauge symmetry that reproduces all the experimentally observed
differential cross sections in nature, including those produced by gravitational interactions to $\mathcal{O}(\frac{v^2}{c^2})$.
QGD also has the ability to fit those diagrams contributions from gravity that may become available in the future such as Higgs-Gluon gravitational scattering.
In addition, QGD has gravitational SU(2) Yang Mills BPS states that match the entropy of black holes and type-II strings.

Pure Yang's gravity\cite{yang} has similar classical equations to those of Einstein's Relativity and includes Schwarzschild-like solutions, 
whose singularity at the horizon, as opposed to black hole solutions, cannot be removed through a coordinate transformation\cite{DV}.
QGD and Yang's gravity are both constructed in terms of the same connection defined over a fiber bundle, However, Yang's gravity 
is not renormalizable because it couples its Yang-Mills field directly to the  energy-momentum tensor. Instead, in QGD, the Yang Mills fields
couple to matter through a $SO(4)\times U(1)$ covariant derivative, in the same manner as the other forces in nature and which preserves renormalization.
Therefore, the couplings in QGD ensure that all terms in the Lagrangian have dimensionality
four or less which guarantees renormalization\cite{wein2}.

String theory is the only other theory, besides QGD, to incorporate gravitational interactions while preserving renormalization.
It has been a successful gravitational theory, because like QGD, scattering processes associated with the $\beta$-function constrain the
metric to be a solution of Einstein's equation. Therefore, like QGD it reproduces all General Relativity tests to $\mathcal{O}(\frac{v^2}{c^2})$.
Like strings, QGD has the necessary BPS states to resolve the black hole entropy problem.
However, unlike QGD, string theory cannot be defined without ambiguity in 4 dimensions and 
neither can it couple to the Standard Model which instead must be reformulated along side gravity. 

The QGD connection defined on the SO(4) fiber bundle can be projected to the connection defined on 
the SO(1,3) tangent bundle. This projection allows
the generators of the SO(4) gauge fields to be defined in terms of Dirac gamma matrices
\begin{equation}
T^{ab}=-\frac{i}{4}[\gamma^a,\gamma^b],\label{SO4g}
\end{equation}
and therefore these generators 
do not commute with the spacetime Dirac $\gamma_\mu$ matrices. This non-commutativity contrasts QGD from the other gauge theories in the Standard Model because
it introduces terms proportional to $[\gamma_\mu,T^{ab}]$ which couple to fermionic matter fields with arbitrary couplings.
The map from the SO(4) gravitational gauge field, $\omega_\mu^{ab}$, to the spin-connection defined in\cite{wein} motivate calling $\omega_\mu^{ab}$
the connecton.

In order to test QGD, we must reproduce the experimental evidence supporting General Relativity. As discussed at the end of section 4, 
this also suffices to show that QGD will reproduce all effects of General Relativity to $\mathcal{O}(\frac{v^2}{c^2})$. 
The procedure used
to fit QED to experimental evidence, see for example\cite{gribov}, transfers well
to QGD and permits the fitting of all the QGD parameters to experimental observation. 
The  T-matrix calculation in\cite{gribov} associated with the tree-level $e^-e^-\to e^-e^-$ scattering process produces a differential cross section;
comparison of that differential cross section with the one 
obtained from the Schroedinger equation with a Coulomb potential allows the fitting of the electron coupling found in QED to the experimentally 
measured charge used in the Coulomb potential. 
This procedure extends to the quantized but non-renormalizable General Relativity theory.  In\cite{don}
the tree-level scattering process for two Bosons, $b\,b'\to b\,b'$, produces a differential scattering matrix which 
fits the Schroedinger equation with a Newton potential while the  1-loop scattering correction fits the differential cross section 
produced by the Schroedinger equation with the leading classical post Newtonian correction to the potential.
The strategy implemented in\cite{gribov} and\cite{don} was also successfully implemented to the gravitational sector  in string theory
\cite{am1}-\cite{am4}. Thus, matching the  model's differential cross sections to experimental/expected differential cross sections allows us to 
test if a model correctly describes nature. 

Here we follow the same strategy as that used in\cite{gribov}-\cite{am4} to show that QGD coupled to the Standard Model
fits all the known experimental differential cross sections, including those gravitational in nature.
Section 2 details the symmetries and action derived for a single matter field and the U(1) electromagnetic field, both coupled to gravity.  
In section 3 we calculate the 
propagators and vertices of the model presented in section 2. In section 4 we calculate T-matrix elements to produce differential cross sections
which are then fitted to the differential cross section of the Schroedinger equations with the Coulomb, the Newtonian and post Newtonian potentials as
well as the differential cross section for the deflection of light by a point particle in the small angle approximation found in General Relativity.
Section 5 presents the full QGD coupling to the Standard Model while section 6 fits the differential cross sections 
to experimentally observed differential cross sections. Section 7 moves outside the realm of experimental observation and shows that QGD can include the expected
gravitational interactions of the SU(2) and SU(3) gauge Bosons as well as the Higgs doublet. 
In section 8 we consider the
isomorphic $SU(2)\times SU(2)$ Yang Mills representation of QGD, which when $N=2$ supersymmetrized has BPS states whose entropy corresponds to that of a black hole, 
and that can bridge to type-II strings\cite{vafa}.
The conclusions appear in section 9.

\section{Symmetries and Actions}

The fundamental geometrical object of General Relativity, the connection on the tangent bundle, does not require a metric to describe a
particle's geodesics.  Consistent with General Relativity and in the spirit of\cite{yang}, 
QGD chooses a connection over the metric as the fundamental field. QGD does not use a yet to be detected graviton field or spin-2 particle
and it uses instead a constant metric void of any dynamics.  
Instead, connections defined on the SO(4) fiber bundle mediate 
all gravitational interactions; 
these SO(4) gauge fields alone mediates all the gravitational interactions between particles.  
These fields suffice to obtain all the differential cross sections expected from General Relativity for a single matter-photon system.
More importantly, this construction does not have any experimental impediments because we have yet to detect the graviton or any other spin-2 particle.

We present the Lagrangian
\begin{eqnarray}
L&=&L_{gauge}+L_f\label{lagtodo}\\
L_{gauge}&=&\frac{1}{4}F_{\mu\nu}F^{\mu\nu}+\frac{1}{4}\Omega_{\mu\nu}^{ab}\Omega_{\mu\nu}^{ab}\label{lagB}\\
L_f&=& \frac{i}{2}\bar{\psi}\{\gamma^\mu, D_\mu\}\psi+\frac{g'}{g}\bar{\psi}[\gamma^\mu, D_\mu]\psi.\label{lagF}
\end{eqnarray}
We use the  following conventions. 
The diagonal and flat space-time metric, $\eta^{\mu\nu}$, does not carry any dynamics and has signature $(+,-,-,-)$.
$\epsilon^{\mu\nu\rho\sigma}$ represents the Levi-Civita tensor in 4 dimensions.  The vielbeins  $e^a_\rho$ connect space-time coordinates (Greek indices) 
with the SO(4) fiber bundle coordinates (Latin indices). In particular $e^a_\rho e^{b\rho}= \eta^{ab}$ and $e^{a\rho} e_a^\sigma= \eta^{\rho\sigma}$. The metric
$\eta^{ab}$, defined on the fiber bundle, has signature $(+,+,+,+)$.

We define the field strengths appearing in (\ref{lagB}) 
\begin{eqnarray}
F_{\mu\nu}&=&\frac{1}{\beta}(\partial_\mu  A_\nu-\partial_\nu  A_\mu) \label{beq3} \\
\Omega_{\mu\nu}^{ab}&=&(\partial_\mu\omega_\nu^{ab}+\frac{\alpha}{\beta}\epsilon_\nu^{ab\rho}\partial_\mu A_\rho)-(\partial_\nu\omega_\mu^{ab}+\frac{\alpha}{\beta}\epsilon_\mu^{ab\rho}\partial_\nu A_\rho)\nonumber\\
& &-g C^{ab}_{\;\;\;cdef}(\omega_\mu^{cd}+\frac{\alpha}{\beta}\epsilon_\mu^{cd\rho} A_\rho)(\omega_\nu^{ef}+\frac{\alpha}{\beta}\epsilon_\nu^{ef\sigma} A_\sigma).\nonumber\\
& & \label{beq1}
\end{eqnarray}
$A_\mu$ represents the U(1) electromagnetic gauge field and the SO(4) gauge field $\omega_{\mu}^{ab}$, the connecton, mediates of gravitational interactions. 
The couplings $e$, $g$, $g'$ and $\alpha$ denote coupling constants and $\beta=\sqrt{1+6\alpha^2}$.
With $\tilde\omega_{\mu ab}$ defined as a copy of a textbook SO(N=4) 
gauge field necessarily antisymmetric in indices $a$ and $b$, then
\begin{equation}
\omega_{\mu ab}=\tilde{\omega}_{\mu ab}+e^{c}_{\mu} e^\nu_{a} \tilde{\omega}_{\nu cb},\label{prop1}
\end{equation}
so that the indispensable property
\begin{equation}
\omega_{\mu ab}e^{a}_\nu e^{b}_\rho\epsilon^{\mu \nu\rho\sigma}=0\label{prop1a}
\end{equation}
holds.

The covariant derivative operator $D_\mu$ defined in (\ref{lagF}) acts on a 4 dimensional Dirac spinor field $\psi$
\begin{equation}
D_\mu\psi=(
\partial_\mu + \frac{i e}{\beta} A_\mu+ i g ( \omega_\mu^{ ab}+\frac{\alpha}{\beta} \epsilon_\mu^{ab\nu}A_\nu) T_{ ab})\psi
\label{eq1a}
\end{equation}
and depends on the SO(4) generators $T^{ab}$ defined in (\ref{SO4g}).

The U(1) local gauge transformation $\Gamma=exp(i\lambda)$, with $\lambda$ a scalar function, acts on the following fields
\begin{eqnarray}
\psi'&=&\Gamma\psi\\
 A'_\mu&=& A_\mu-\frac{\beta}{e}\partial_\mu\lambda \label{eqx0}\\
\omega_\mu^{'ab}&=& \omega_\mu^{ab}+\frac{\alpha}{e}\epsilon_\mu^{ab\nu}\partial_\nu\lambda\label{eqx11}
\end{eqnarray}
which also transforms the covariant derivative in the standard way
\begin{equation}
D_\mu'=D_\mu-i\partial_\mu\lambda,
\end{equation}
but leaves (\ref{lagtodo}) unchanged.

The SO(4) local gauge transformations with $\hat{\Gamma}=exp(i\Lambda^{ab}T_{ab})$ and $\Lambda^{ab}$ a 2-Tensor, act on the following fields
\begin{eqnarray}
\psi'&=&\hat\Gamma\psi\\
A'_\mu&=& A_\mu\label{eqx2}\\
\omega_\mu^{'ab}&=&\omega_\mu^{ab}-\frac{1}{g}\partial_\mu\Lambda^{ab}   + C^{ab}_{\;\;\;cdef}\Lambda^{cd}(
\omega_\mu^{ef}+\frac{\alpha}{\beta}\epsilon_\mu^{ef\nu}A_\nu)\label{eqx21}
\end{eqnarray}
which also transforms the covariant derivative in the standard way
\begin{equation}
D_\mu'=D_\mu-iT_{ab}\bigg[\partial_\mu\Lambda^{ab}-C^{ab}_{\;\;\;cdef}\Lambda^{cd}(
\omega_\mu^{ef}+\frac{\alpha}{\beta}\epsilon_\mu^{ef\nu}A_\nu)\bigg],
\end{equation}
but also leaves (\ref{lagtodo}) unchanged.

The SO(1,3) local gauge transformations acting on the Fermions have the same effect as the SO(4) transformation because
\begin{equation}
\hat{\Gamma}=e^{i\Lambda^{ab}T_{ab}}= e^{-i \Lambda^{ab}T_{cd} e^\mu_a e^\nu_b e_\mu^c e_\nu^d}=e^{i\Lambda^{\mu\nu}T_{\mu\nu}}.
\end{equation}
However, the SO(1,3) gauge transformations, can not act on the pure Yang-Mills Lagrangian because the Cartan-Killing metric for that gauge group, given 
by the trace over two group elements, is not positive definite which yields states not bounded from below\cite{wein2}.  This property therefore
distinguishes theories defined over the tangent bundle with local SO(1,3) symmetry from those defined over the fiber bundle with local SO(4) symmetry. The
former cannot be used to define Yang-Mills theories even when the fiber bundles can be projected to the tangent bundle. For the avoidance
of doubt, gravitational theories like spin-gravity can be defined over the tangent bundle only because, as a result of 
the different construction from Yang-Mills, their spectrum does not become unbounded from below when the Cartan-Killing metric is not positive definite.

The Lorenz gauge fixings
\begin{eqnarray}
\partial_\mu  A^\mu&=&0\\
\partial_\mu \omega^{\mu ab}&=&0
\end{eqnarray}
and property (\ref{prop1}) simplify the quadratic component in (\ref{lagB}) to 
\begin{equation}
L_{gauge}^{(2)}=\frac{1}{2}\partial_\mu  A^\nu\partial^\mu A_\nu
+\frac{1}{2}\partial_\mu\omega^{\nu ab}\partial^\mu\omega_{\nu ab}.\label{lag12}
\end{equation}

The structure constant defined in\cite{wein}
\begin{equation}
C^{abcdef}=-\eta^{af}\eta^{ce}\eta^{bd}+\eta^{ad}\eta^{ce}\eta^{bf}-\eta^{ae}\eta^{cb}\eta^{df}+\eta^{ac}\eta^{eb}\eta^{df}
\end{equation}
simplifies the relevant terms of the cubic terms in (\ref{lagB}) to 
\begin{eqnarray}
L_{gauge}^{(3)}&=& 2 g (\partial_\mu\omega_\nu^{ab}-\partial_\nu\omega_\mu^{ab}) \omega^{\mu }_{\;\;c a}\; \omega^{\nu c }_{\;\;\;b}
\nonumber\\
& &
-4g\frac{\alpha^2}{\beta^2} (A_\mu \partial_\rho A_\nu e^\nu_a e^\rho_b \omega^{\mu a b }+ \partial_\nu A_\mu A_\rho e^\mu_a e^\rho_b\omega^{\nu a b})
\label{lag13}.
\end{eqnarray}
The terms in (\ref{lag13}) take the form $\omega\omega\partial \omega$ and  $\partial A A \omega$.
In the low energy limit the photon momentum, $r$, and the connecton momentum, $k$, satisfy
\begin{equation}
r>>k
\end{equation}
or 
equivalently 
\begin{equation}
\partial A>>\partial \omega.
\end{equation} This condition accounts for the omission of all other terms. 
For example, for matter-photon scattering
the vertex generated by $A\omega\partial\omega$ would only participate at 1-loop level, would necessarily appear twice and thus takes values proportional
to the square of the connecton momentum. Instead, the vertices produced by the second term in (\ref{lag13}) would appear first at tree-level
and then at 1-loop level where it would contribute terms proportional to the photon momentum and its square respectively.

Dimensional analysis of General Relativity shows that it does not renormalize properly. 
However, it follows from (\ref{eq1a}) that the connecton has the same dimensionality as the U(1) gauge field.  Therefore, the terms in the Lagrangian
(\ref{lag12}) and (\ref{lag13})  involve products of fields and their derivatives with total dimensionality four or less.  
Renormalization then follows as shown in\cite{wein2}.

Since $[\gamma_\mu,T_{ab}]\ne0$, we use 
the relation
\begin{equation}
\gamma^\mu\gamma^\nu\gamma^\lambda = \eta^{\mu\nu}\gamma^\lambda + \eta^{\nu\lambda}\gamma^\mu - \eta^{\mu\lambda}\gamma^\nu - i\epsilon^{\sigma\mu\nu\lambda}\gamma_\sigma\gamma^5, 
\end{equation}
to express
\begin{eqnarray}
\frac{i}{2}\bar{\psi}\{\gamma^\mu, D_\mu\}\psi&=&i\bar\psi(\slashed\partial+  \frac{i e}{\beta}\slashed A+ 3 i g\frac{ \alpha}{\beta} A^\mu\gamma^5\gamma_\mu)\psi ,\label{lf1}\\
i\bar{\psi}[\gamma^\mu, D_\mu]\psi&=&g \bar\psi\bar\omega_{\;\;\;\;\mu}^{\mu a}\gamma_a\psi.\label{lf2}
\end{eqnarray}
Note that (\ref{lf2}) does not depend on $\partial_\mu$.  This means that gauge invariance constrains (\ref{lf1}) but not (\ref{lf2}). Thus
the rescaling of (\ref{lf1}) by a constant redefines the coupling. However,
  the redefinition guarantees gauge invariance only after a subsequent 
redefinition of the spinor $\psi$. 
On the other hand, a recalling by a constant, even a complex one, of (\ref{lf2}) does not affect gauge invariance.
This contrasts with SU(2) of SU(3) couplings where the Dirac matrices commute with their generators and therefore
(\ref{lf2}) necessarily vanishes.

The relations (\ref{lf1}) and (\ref{lf2}) simplify the  Lagrangian (\ref{lagF}) to 
\begin{equation}
L_{f}=\bar\psi (i(\slashed\partial+  \frac{i e}{\beta}\slashed A+ 3 i g \frac{\alpha}{\beta} A^\mu\gamma^5\gamma_\mu+g' \omega_{\;\;\;\;\mu}^{\mu a}\gamma_a)-m)\psi. 
\label{lag2}
\end{equation}

Here the real coupling constants $e$, $g$, $\alpha$ and $g'$ adjust the theory to observation.  In particular 
$e$ represents the electric charge and 
gauge invariance constrains $g$ and coincides with the coupling constant found in the purely bosonic sector (\ref{lag13}).
Instead, gauge invariance does not constrain $g'$ 
which can  take any value, real, purely imaginary or complex. 

Again, the terms in the Lagrangian
(\ref{lag2})  involve products of fields and their derivatives with total dimensionality four or less.  
Therefore, the terms in the Lagrangians (\ref{lag12}), (\ref{lag13})
and (\ref{lag2}), and thus Lagrangian (\ref{lagtodo}), involve products of fields and their derivatives with total dimensionality four or less.  
Renormalization then follows as shown in\cite{wein2}.

The Lagrangian $L_f$ has an complex coupling constant $g'$ while $L^{(3)}_{gauge}$ has a real coupling constant $g$.  Then,
self energy corrections can have imaginary contributions 
and therefore it would be tempting to assume that the probability density determined by the wavefunction decreases over time and unitarity is lost. 
This is not the case.

We choose a pure real $g$ and a pure imaginary $g'$  and will show when fitting theory to experiment in the subsequent sections that 
\begin{equation}
g\sim g' \sim \mathcal{O}(\sqrt{G}). 
\end{equation}
where $G$ represents the gravitational constant. 

The self-energy contributions to $\mathcal{O}(G^2)$ for both Bosons $A_\mu$ and $\omega_\mu^{ab}$ come from purely bosonic diagrams and those with fermionic loops.  
We construct the purely bosonic diagrams with vertex (c) and vertex (d) defined in section \ref{pv1}. Therefore, the diagrams 
depend only on the real valued coupling constant $g$
\begin{equation}
\Sigma_A^{(G)}\propto g^2,\;\;\;
\Sigma_A^{(G^2)}\propto g^4,\;\;\;
\end{equation}
\begin{equation}
\Sigma_\omega^{(G)}\propto g^2,\;\;\;
\Sigma_\omega^{(G^2)}\propto g^4.
\end{equation}
Thus, those diagrams always contribute real amounts to the self energies $\Sigma_A^{(G)}$, $\Sigma_\omega^{(G)}$, $\Sigma_A^{(G^2)}$ and $\Sigma_\omega^{(G^2)}$.
As we show below, real contributions to the self energy do not affect unitarity.

The diagrams involving fermionic loops always involve an even number of vertices (a) or an even number of vertices (b) defined in section \ref{pv1}.  Therefore,
corrections to $\mathcal{O}(G^2)$ to the self energies of both $A_\mu$ and $\omega_\mu^{ab}$ always contribute even powers of $g'$  
\begin{equation}
\Sigma_A^{(G)}\propto g^{'2},\;\;\;
\Sigma_A^{(G^2)}\propto a_1 g^{'4}+a_2 g^{'2}g^2,\;\;\;
\end{equation}
\begin{equation}
\Sigma_\omega^{(G)}\propto g^{'2},\;\;\;
\Sigma_\omega^{(G^2)}\propto a_1 g^{'4}+a_2 g^{'2}g^2
\end{equation}
for some real constants $a_1$ and $a_2$.
Even though the coupling $g'$ is imaginary,
all corrections to the self energies for both Bosons $A_\mu$ and $\omega_\mu^{ab}$ contribute real amounts to $\mathcal{O}(G^2)$, 
Even more, because Fermions always appear in loops regardless of the order of perturbation, the self energy corrections to the Bosons $A_\mu$ and $\omega_\mu^{ab}$ 
always remain real.

On the other hand, the contributions to the self energies of the Fermion contribute  real amounts to $\mathcal{O}(G)$ but complex amounts to $\mathcal{O}(G^2)$.
The contribution to the Fermion self energy to $\mathcal{O}(G)$ comes from diagrams constructed from 
a pair of vertices (a) or a pair of vertices (b). Therefore
\begin{equation}
\Sigma_f^{(G)}\propto b_1 g^{'2}+ b_2 g^2\;\;\;
\end{equation}
for some real constants $b_1$ and $b_2$.
The first imaginary contribution comes from the diagram involving one vertex (c) and three vertices (a). Therefore
\begin{equation}
\Sigma_f^{(G^2)}\propto g^{'3} g \sim iG^2.
\end{equation}
Therefore, the leading imaginary correction to the Fermion self energy contributes
at most of  $\mathcal{O}(G^2)$. 

The time evolution of the wavefunction for the Fermion  
\begin{equation}
|\psi(t_{out})>=e^{-i H (t_{out}-t_{in})}|\psi(t_{in})>=e^{-i (m+ Re(\Sigma_f)+i Im (\Sigma_f)) (t_{out}-t_{in})}|\psi(t_{in})>,
\end{equation}
where $|\psi(t_{in})>$ describes the wavefunction at the emitter and $|\psi(t_{out})>$ represents the wavefunction at the detector,
implies that the probability density at time $t_{out}$, $<\psi(t_{out})|\psi(t_{out})>$, decays as 
\begin{eqnarray}
<\psi(t_{out})|\psi(t_{out})>&=&e^{-2 Im(\Sigma_f (t_{out}-t_{in})}<\psi(t_{in})|\psi(t_{in})>\nonumber\\
&=&e^{-2 Im(\Sigma_f^{(G^2)}) (t_{out}-t_{in})}<\psi(t_{in})|\psi(t_{in})>  +\mathcal{O}(G^3).\;\;\;\label{flux}
\end{eqnarray}
When $<\psi(t_{out})|\psi(t_{out})> < <\psi(t_{in})|\psi(t_{in})>$ 
particles are lost as they travel between emitter and detector. This experimentally unobserved
reduction in flux prevents large imaginary values of $\Sigma_p$.   However, when $Im(\Sigma_f)$ satisfies
\begin{equation}
Im(\Sigma_f) (t_{out}-t_{in})<<<1,
\end{equation} 
then, the self energy $\Sigma_f$ can take on imaginary values without loss of unitarity.  
Since the leading imaginary correction to the self energy of the Fermion contributes at most  $\mathcal{O}(G^2)$, then, 
\begin{equation}
-2\Sigma_f^{(G^2)} (t_{out}-t_{in})<<<1
\end{equation}
 whenever $t_{out}-t_{in}<10^{34}$ years.  Bearing in mind that the universe has existed less than $10^{11}$ years and that
estimates put the the proton's half life at $10^{32}$ years,
QGD can take imaginary couplings without affecting unitarity. This contrasts with other theories where the coupling constants are
much larger than $g'$ leading to a loss of unitarity if not of gauge invariance as well since $g'\ne g$. This conclusion is
further supported by the absence of diagrams
contributing to $\mathcal{O}(G^2)$ to the self energies constructed with vertex (a) and vertices involving the electromagnetic field.

\section{Propagators and Vertices}\label{pv1}

The Lagrangian densities (\ref{lag12}) and (\ref{lag2}) yield the following propagators
\begin{center}
\begin{tabular}{|p{5cm}|p{4cm}|}
\hline
\begin{center}
\begin{fmffile}{gluon}
\begin{fmfgraph*}(70,15)
\fmfleft{in}
\fmfright{out}
\fmflabel{$_\mu^{ab}$}{in}
\fmflabel{$_\nu^{cd}$}{out}
\fmf{dbl_curly,label=$k$,label.dist=10}{in,out}
\end{fmfgraph*}
\end{fmffile}
\end{center}
\begin{center}Propagator (a)\end{center}
 &\vspace{0cm} \Large $-\frac{i g^{\mu\nu}\eta^{ab}\eta^{cd}}{k^2+i\epsilon}$ \vspace{0.5cm} \\ \hline
\begin{center}
\begin{fmffile}{fermion}
\begin{fmfgraph*}(70,15)
\fmfleft{in}
\fmfright{out}
\fmflabel{$\alpha$}{in}
\fmflabel{$\beta$}{out}
\fmf{fermion,label=$p$}{in,out}
\end{fmfgraph*}
\end{fmffile}
\end{center}
\begin{center}Propagator (b)\end{center}
& \vspace{0cm}\Large $\frac{i\delta^{\alpha\beta}}{\slashed{p}-m+i\epsilon}$\vspace{0.5cm}  \\ \hline
\begin{center}
\begin{fmffile}{photon}
\begin{fmfgraph*}(70,15)
\fmfleft{in}
\fmflabel{$\mu$}{in}
\fmflabel{$\nu$}{out}
\fmfright{out}
\fmf{photon,label=$r$}{in,out}
\end{fmfgraph*}
\end{fmffile}
\end{center}
\begin{center}Propagator (c)\end{center}
 &\vspace{0cm}\Large $-\frac{ig^{\mu\nu}}{r^2+i\epsilon}$\vspace{0.5cm}\\
\hline
\end{tabular}
\end{center}
For the avoidance of doubt, in the limit considered here
\begin{equation}
m>>\partial A>>\partial\omega
\end{equation}
 or equivalently $m>>r>>k$.
The Lagrangian densities (\ref{lag13}) and (\ref{lag2}) generate the following vertices:
\begin{center}
\begin{tabular}{ |p{4.5cm}| p{8.5cm}| }
\hline
\begin{center}
\begin{fmffile}{vertex1}
\begin{fmfgraph*}(100,65)
\fmfleft{i1,i2}
\fmfright{o1}
\fmf{fermion,tension=0.5}{i1,v1}
\fmf{fermion,tension=0.5}{v1,i2}
\fmf{dbl_curly,tension=0.5}{v1,o1}
\fmflabel{$\alpha$}{i1}
\fmflabel{$\beta$}{i2}
\fmflabel{$_\mu^{ab}$}{o1}
\end{fmfgraph*}
\end{fmffile}
\end{center}
\begin{center}Vertex (a)\end{center}
&\vspace{1cm}\large $ -g' \; e^{\mu a} e^{\nu b} \gamma_\nu $ \\ \hline
\begin{center}
\begin{fmffile}{vertex2}
\begin{fmfgraph*}(100,65)
\fmfleft{i1,i2}
\fmfright{o1}
\fmf{fermion,tension=0.5}{i1,v1}
\fmf{fermion,tension=0.5}{v1,i2}
\fmf{photon,tension=0.5}{v1,o1}
\fmflabel{$\alpha$}{i1}
\fmflabel{$\beta$}{i2}
\fmflabel{$_\mu$}{o1}
\end{fmfgraph*}
\end{fmffile}
\end{center}
\begin{center}Vertex (b)\end{center}
&\vspace{1cm} \large $ -\frac{i }{\beta} (e \gamma_\mu+3g\alpha \gamma_5\gamma_\mu)  $ \\ \hline
\begin{center}
\begin{fmffile}{vertex3}
\begin{fmfgraph*}(100,70)
\fmfleft{i1,i2}
\fmfright{o1}
\fmf{dbl_curly,tension=1.5,label=$p$,label.dist=-15}{i1,v1}
\fmf{dbl_curly,tension=1.5,label=$q$,label.dist=-15}{v1,i2}
\fmf{dbl_curly,tension=1.5,label=$r$,label.dist=-15}{v1,o1}
\fmflabel{$_\mu^{ab}$}{i1}
\fmflabel{$_\nu^{cd}$}{i2}
\fmflabel{$_\rho^{ed}$}{o1}
\end{fmfgraph*}
\end{fmffile}
\end{center}
\begin{center}Vertex (c)\end{center}
 &\vspace{1cm} \small$g(h^{af}h^{ce}h^{bd}+h^{ad}h^{ce}h^{bf}-h^{ae}h^{cb}h^{df}+h^{ac}h^{eb}h^{df})\cdot$
 $\cdot((p-q)^\rho g^{\mu\nu}+(q-r)^\mu g^{\nu\rho}+(r-p)^\nu g^{\mu\rho})$  \\ \hline
\begin{center}
\begin{fmffile}{vertex4}
\begin{fmfgraph*}(100,65)
\fmfleft{i1,i2}
\fmfright{o1}
\fmf{photon,tension=1.5,label=$r$}{i1,v1}
\fmf{photon,tension=1.5,label=$r'$}{v1,i2}
\fmf{dbl_curly,tension=1.5,label=$q$,label.dist=-15}{v1,o1}
\fmflabel{$\beta$}{i2}
\fmflabel{$\alpha$}{i1}
\fmflabel{$_\mu^{ab}$}{o1}
\end{fmfgraph*}
\end{fmffile}
\end{center}
\begin{center}Vertex (d)\end{center}
&\vspace{1cm}\large $ g\frac{4\alpha^2}{\beta^2}(g_{\mu\alpha}e^a_\beta r^{'b}+ g_{\mu\beta}e^a_\alpha r^{b} 
+e^a_\alpha e^b_\beta r_\mu   +e^a_\beta e^b_\alpha r'_\mu) $ \\ \hline
\end{tabular}
\end{center}

In the following section we relate the tree and 1-loop level diagrams  to physical processes in order to determine the four parameters
in QGD.

\section{Tree and 1-loop Diagrams and Their Relationship to Observed Quantities} 

In this section we calculate 4 scattering amplitudes: the tree-level photon exchange between two particles with equal mass, the tree-level connecton exchange and the 1-loop connecton exchange between two particles with equal mass and the tree-level connecton exchange between a photon and a particle with mass.
These scattering amplitudes 
yield a total of 4 differential cross sections; we equate the first 3 to the differential cross section for a Schroedinger equation with 3 different potentials: the
Coulomb potential, the Newton potential and the post-Newtonian potential while the last cross section matches the cross section for 
gravitational lensing of a photon by a massive point particle. The Lagrangian provides 4 coupling constants: $e$, $g$, $g'$ and $\alpha$.  Therefore
by the end of this section, the 4 coupling constants appearing in the differential cross sections calculated from QGD match those of the 
Schroedinger equation as well as that obtained from gravitational lensing in General Relativity.

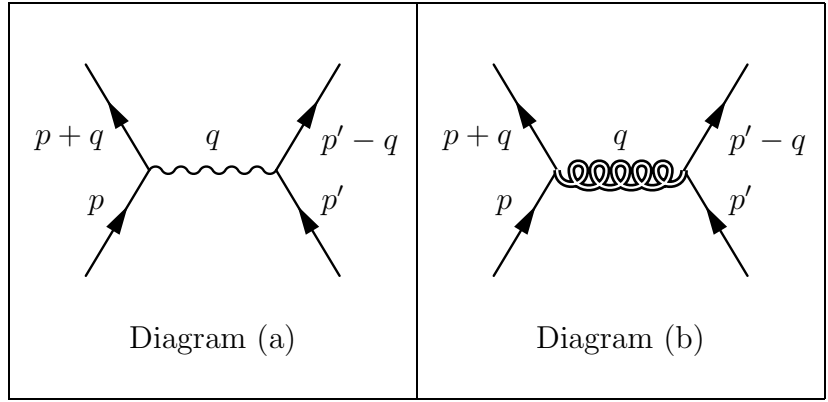
\begin{center}
\begin{tabular}{ |p{5cm}| p{5cm}| }

\hline

\begin{center}
\begin{fmffile}{dia2}
\begin{fmfgraph*}(120,80)
\fmfleft{i1,i2}
\fmfright{i3,i4}
\fmf{fermion,tension=0.2,label=$p$,label.side=left}{i1,v1}
\fmf{fermion,tension=0.2,label=$p+q$,label.side=left}{v1,i2}
\fmf{fermion,tension=0.2,label=$p'$,label.side=right}{i3,v2}
\fmf{fermion,tension=0.2,label=$p'-q$,label.side=right}{v2,i4}
\fmf{photon,tension=0.2,label=$q$,label.dist=-15}{v1,v2}
\end{fmfgraph*}
\end{fmffile}
\end{center}
\begin{center}Diagram (a)\end{center}

&

\begin{center}
\begin{fmffile}{dia1}
\begin{fmfgraph*}(120,80)
\fmfleft{i1,i2}
\fmfright{i3,i4}
\fmf{fermion,tension=0.2,label=$p$,label.side=left}{i1,v1}
\fmf{fermion,tension=0.2,label=$p+q$,label.side=left}{v1,i2}
\fmf{fermion,tension=0.2,label=$p'$,label.side=right}{i3,v2}
\fmf{fermion,tension=0.2,label=$p'-q$,label.side=right}{v2,i4}
\fmf{dbl_curly,tension=0.2,label=$q$,label.dist=-15}{v1,v2}
\end{fmfgraph*}
\end{fmffile}
\end{center}
\begin{center}Diagram (b)\end{center}
\\
\hline
\end{tabular}
\end{center}

\vspace{1cm}

The contribution from the photon exchange between two mass particles in diagram  (a) produces the T-matrix
\begin{equation}
T_{QED}=4i m^2\frac{ e^2/\beta^2  }{-q^2}+4i e g \frac{3 \alpha}{\beta}\frac{p^\mu J'_{A\mu}+p^{'\mu} J_{A\mu}}{-q^2}   +4i g^2 \frac{9   \alpha^2}{\beta}\frac{ J_A^\mu J'_{A\mu}}{-q^2}\label{tree-a}
\end{equation}
with $J_{A\mu}=\bar\psi\gamma_\mu\gamma_5\psi$ and $J'_{A\mu}=\bar\psi'\gamma_\mu\gamma_5\psi'$.
The property 
\begin{equation}
p^\mu J'_{A\mu}+p^{'\mu}J_{A\mu} \sim \mathcal{O}(m)<<\mathcal{O}(m^2)
\end{equation}
suppresses the second  term in (\ref{tree-a}).
A further limiting to the spinless Schroedinger limit requires summation over the spin of all in-states and out-states so that $<J_{A\mu}>=<J'_{A\mu}>=0$.
Thus, in the non-relativistic quantum limit, the third term in (\ref{tree-a}) vanish exactly reducing (\ref{tree-a}) to  
\begin{equation}
T_{QED}=4i m^2\frac{ e^2/\beta^2 }{-q^2}.
\end{equation}
This T-matrix yields the differential cross section 
\begin{equation}
\frac{d\sigma_{QED}}{d\Omega}=\frac{m^2}{q^4} e^4/(16\pi^2\beta^4),\label{cross-a}
\end{equation}
while the Schroedinger equation with a Coulomb potential, 
\begin{equation}
V_{C}= \frac{e_{exp}^2}{4\pi R},
\end{equation}
with $e_{exp}$, the measured electric charge, and $R$, the distance between the two particles, produces a differential cross section
\begin{equation}
\frac{d\sigma_{C}}{d\Omega}=\frac{m^2}{q^4} e_{exp}^4/16\pi^2\label{coulomb-a}.
\end{equation}
Comparing the differential cross section (\ref{cross-a}) and (\ref{coulomb-a}) yields
\begin{equation}
e_{exp}=\frac{e}{\beta}.\label{eq-a}
\end{equation}

Diagram (b) contributes to the differential scattering cross section of the connecton exchange between two mass particles.  The T-matrix element
\begin{equation}
T_{QGD,N}= -4im^2  \frac{4 g^{'2} }{-q^2} 
\label{tree1}
\end{equation}
produces the quantum gravitodynamic differential cross section
\begin{equation}
\frac{d\sigma_{QGD,N}}{d\Omega}=\frac{m^2}{q^4}\frac{ g^{'4}}{\pi^2}.\label{cross-b}
\end{equation}
The Schroedinger equation with a Newtonian potential,
\begin{equation}
V_{N}=-G \frac{m_{exp}^2}{R},
\end{equation}
with $G$ representing the gravitational constant, $m_{exp}$ representing 
the measured mass and $R$ representing the distance between the two particles, produces a differential cross
section
\begin{equation}
\frac{d\sigma_{N}}{d\Omega}=\frac{m_{exp}^2}{q^4}G^2 m_{exp}^4.  \label{newton-b}
\end{equation}
Comparing the differential cross section (\ref{cross-b}) and (\ref{newton-b}) yields
\begin{equation}
g^{'}=(\pi G)^{1/2}  m_{exp}\label{tree-b}
\end{equation}
and 
\begin{equation}
m=m_{exp}.\label{tree-b2}
\end{equation}

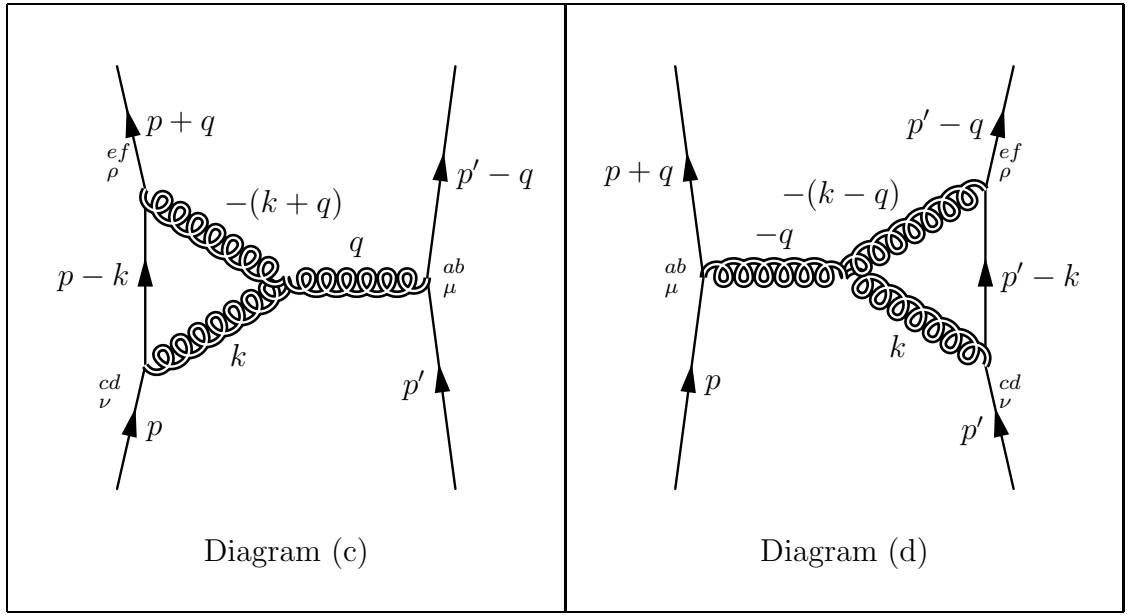
\begin{center}
\begin{tabular}{|p{7cm}|p{7cm}|}
\hline

\begin{center}
\begin{fmffile}{dia3}
\begin{fmfgraph*}(160,160)

\fmfleft{i1,i2}
\fmfright{j1,j2}

\fmf{fermion,label=$p-k$,label.side=left,tension=0.6}{v1,v2}
\fmf{fermion,label=$p+q$,label.side=right}{v2,i2}
\fmf{fermion,label=$p$}{i1,v1}

\fmf{dbl_curly,label=$k$,tension=.2,label.side=right,label.dist=10}{v1,v3}
\fmf{dbl_curly,tension=.2,label=$-(k+q)$,label.side=left}{v2,v3}
\fmf{dbl_curly,label=$q$,tension=.4,label.dist=-15}{v3,v4}
\fmf{fermion,label=$p'$}{j1,v4}
\fmf{fermion,label=$p'-q$}{v4,j2}

\fmflabel{$^{ef}_\rho$}{v2}
\fmflabel{$^{cd}_\nu$ }{v1}
\fmflabel{$^{ab}_\mu$}{v4}

\end{fmfgraph*}
\end{fmffile}
\end{center}
\begin{center}Diagram (c)\end{center}
&
\begin{center}
\begin{fmffile}{dia32}
\begin{fmfgraph*}(160,160)

\fmfleft{j1,j2}
\fmfright{i1,i2}

\fmf{fermion,label=$p'-k$,label.side=right,tension=0.6}{v1,v2}
\fmf{fermion,label=$p'-q$,label.side=left}{v2,i2}
\fmf{fermion,label=$p'$}{i1,v1}

\fmf{dbl_curly,label=$k$,tension=.2,label.side=left}{v1,v3}
\fmf{dbl_curly,tension=.2,label=$-(k-q)$,label.side=right,label.dis=10}{v2,v3}
\fmf{dbl_curly,label=$-q$,tension=.4,label.dist=-20}{v3,v4}
\fmf{fermion,label=$p$}{j1,v4}
\fmf{fermion,label=$p+q$}{v4,j2}
\fmflabel{$^{ef}_\rho$}{v2}
\fmflabel{$^{cd}_\nu$ }{v1}
\fmflabel{$^{ab}_\mu$}{v4}
\end{fmfgraph*}
\end{fmffile}
\end{center}
\begin{center}Diagram (d)\end{center}
\\
\hline
\end{tabular}
\end{center}
\vskip 1.0cm

The post-Newtonian correction comes from diagrams (c) and (d).
All other 1-loop diagrams contribute to either analytic terms, or  quantum corrections of order  $\mathcal{O}(ln(-q^2))$
and do not contribute to the low energy limit considered here\cite{don}.
In addition, diagrams involving 
photons and connectons simultaneously do not contribute because the transformation $\omega_\mu^{'ab}=\omega_\mu^{ab} -\frac{\alpha}{\beta}\epsilon_\mu^{ab\nu}A_\nu $ 
leaves the contribution of the above diagram unchanged while all terms but the first two in (\ref{lag13}) remain present in the action. 
These two diagrams contribute to the quantum gravitodynamic post Newtonian T-matrix
\begin{equation}
T_{QGD,E}= -4im^2\frac{27 g \pi^{3/2} G^{3/2} m^2}{16 \sqrt{-q^2}}. \label{t-cd}
\end{equation}
This result as those below require the following steps.  First, we focus only 
on the ``electric" form factor which means that 
\begin{equation}
\bar\psi\gamma_\mu\gamma_\nu\psi= 2mg_{\mu\nu}.
\end{equation}
Equivalently, we disregard terms proportional to $[\gamma_\mu,\gamma_\nu]$ 
that contribute to the ``magnetic" form factor.
Second, the on-shell external momenta imply the relations
\begin{eqnarray}
p\cdot q&=&-\frac{1}{2}q^2\\
p'\cdot q&=&\frac{1}{2}q^2.
\end{eqnarray}
Third, we suppress  terms of $\mathcal{O}(q^4)$ using  the low energy limit property $q<<p$.
Fourth, we use the approximation $\bar\psi\gamma_\mu\psi=2p_\mu$ and $\bar\psi'\gamma_\mu\psi'=2p'_\mu$.
Finally, we use the appendix in\cite{don} with the expressions for the several Feynman integrals.

The T-matrix (\ref{t-cd}) produces the 
differential scattering cross section 
\begin{equation}
\frac{d\sigma_{QGD,E}}{d\Omega}=\frac{m^2}{|q|^2} \bigg(\frac{27}{16}\bigg)^2 \pi^3\frac{g^2 G^{3} m^4}{16\pi^2}.
\label{cross-cd}
\end{equation}
The Schroedinger equation with a post Newtonian correction to the gravitational potential (see
equation (39) of\cite{don} ),
\begin{equation}
V_{E}=-2 a \frac{  G^2 m_{exp}^3}{R^2}, 
\end{equation}
where the constant $a$ depends on the post-Newtonian expansion,  produces, after using (\ref{tree-b}) and  (\ref{tree-b2}),
the differential scattering cross section 
\begin{equation}
\frac{d\sigma_{E}}{d\Omega}=\frac{m^2}{|q|^2}  4\pi^2 a^2 G^4 m^6.\label{einstein-cd}
\end{equation}

Comparing the differential cross section (\ref{cross-cd}) and (\ref{einstein-cd}) yields
\begin{equation}
g=\frac{128}{27}a \sqrt{\pi G}m. \label{eq-cd}
\end{equation}

Therefore, by fixing $e$, $g$ and $g'$ through equations (\ref{eq-a}), (\ref{tree-b}) and (\ref{eq-cd}), QGD reproduces the low
energy limit of General Relativity for the matter sector.
Therefore,  the SO(4) Yang Mills theory coupled to matter and an inert metric
is a valid renormalizable relativistic quantum field theory to describe gravity's experimental observations.

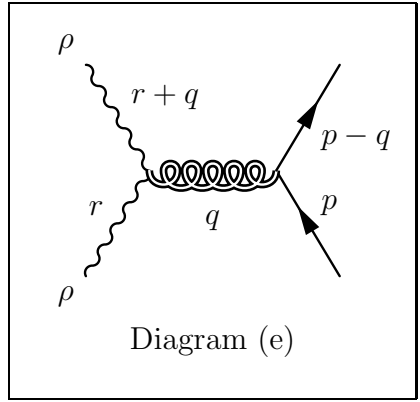
\begin{center}
\begin{tabular}{ |p{5cm}| }
\hline
\begin{center}
\begin{fmffile}{dia6}
\begin{fmfgraph*}(120,80)
\fmfleft{i1,i2}
\fmfright{i3,i4}
\fmf{photon,tension=0.2,label=$r$,label.side=left}{i1,v1}
\fmf{photon,tension=0.2,label=$r+q$,label.side=right}{v1,i2}
\fmf{dbl_curly,tension=.2,label=$q$,label.dist=15}{v1,v2}
\fmf{fermion,tension=0.2,label=$p$,label.side=right}{i3,v2}
\fmf{fermion,tension=0.2,label=$p-q$,label.side=right}{v2,i4}
\fmflabel{$\rho$}{i1}
\fmflabel{$\rho$}{i2}
\end{fmfgraph*}
\end{fmffile}
\end{center}
\begin{center}Diagram (e)\end{center}
\\
\hline
\end{tabular}
\end{center}
Diagram (e) describes the scattering of a Fermion $f$ with a photon $\gamma$ that produces the T-matrix 
\begin{equation}
T_{f\,\gamma\to f\,\gamma}=2im\frac{40 g'g\alpha^2}{\beta^2}  \frac{ E_{\gamma}}{-q^2}, \label{saa0}
\end{equation}
which in turn, 
in the small angle approximation where $q=r sin(\theta/2)$, 
yields the differential cross section 
\begin{equation}
\frac{d\sigma_{f\,\gamma\to f\,\gamma}}{d\Omega}=\frac{g^2g^{'2}}{64\pi^2 }\bigg(\frac{40 \alpha^2}{\beta^2}\bigg)^2 \frac{E_{\gamma}^2}{(r^2 sin^2(\theta/2))^2},\label{saa1}
\end{equation}
where the photon energy, $E_{\gamma}$, satisfies 
\begin{equation}
E_{\gamma}<<m\label{esm1}.
\end{equation} 
Derivation of (\ref{saa1})  requires
the photon on-shell conditions, 
\begin{eqnarray}
r\cdot q&=&\frac{q^2}{2}\\
r'\cdot q&=&-\frac{q^2}{2},
\end{eqnarray}
along with the steps used for the matter-matter scattering above.
Substitution of (\ref{tree-b}) and (\ref{eq-cd}) into (\ref{saa1}) yields
\begin{equation}
\frac{d\sigma_{p\,\gamma\to p\,\gamma}}{d\Omega}=\bigg(\frac{16}{27}\bigg)^2\bigg(\frac{40\alpha^2}{\beta^2}\bigg)^2 a^2 G^2 m^2 \frac{m^2E_{\gamma}^2}{(r^2 sin^2(\theta/2))^2}.\label{saa2}
\end{equation}
In the low energy limit, equation (\ref{saa2}) describes the Rutherford scattering of a mass $m$ projectile off of a
mass $E_{\gamma}$ target. 

In the small angle deflection approximation 
and after setting the speed of light to 1, the gravitational lensing of a photon by a point particle of mass $m$ in General Relativity 
relates the impact parameter $b_{L}$ with the deflection angle $\theta$\footnote{See\cite{mene} for an overview of that calculation.}
\begin{equation}
b_{L}=\frac{4 Gm }{\theta}=2Gm \,cot(\theta/2). 
\end{equation}

The relation between the impact parameter, $b_L$, and the differential cross section,
\begin{equation}
\frac{d\sigma_{L}}{d\Omega} = \frac{b_L}{\sin{\theta}} \left|\frac{db_L}{d\theta}\right|,
\end{equation}
simplifies the latter to 
\begin{equation}
\frac{d\sigma_{L}}{d\Omega} =\frac{G^2m^2}{sin^4(\theta/2)}, 
\end{equation}
or equivalently
\begin{equation}
\frac{d\sigma_{L}}{d\Omega} =G^2 E_{\gamma}^2\frac{m^2 E_{\gamma}^2}{(r^2sin^2(\theta/2))^2} 
\label{saa5}
\end{equation}
Equation (\ref{saa5}) describes the Rutherford  scattering of mass $E_{\gamma}$ projectile off of a mass $m$ target. The exchange $m \leftrightarrow E_{\gamma}$
modifies this differential cross section to
\begin{equation}
\bigg(\frac{d\sigma_{L}}{d\Omega}\bigg)_{m \leftrightarrow E_{\gamma}}=
G^2 m^2\frac{m^2 E_{\gamma}^2}{(r^2sin^2(\theta/2))^2} 
\label{saa3}
\end{equation}
which then describes the scattering
of a mass $m$ particle of a photon target with energy $E_{\gamma}$. 

Equating (\ref{saa2}) with (\ref{saa3}) requires that 
\begin{equation}
\frac{16}{27}\frac{40\alpha^2}{1+6\alpha^2} a =1,
\label{tree-light}
\end{equation}
which determines $\alpha$ as a function of the known constant $a$.

With (\ref{tree-light}) satisfied, Fourier transform in the Born approximation of (\ref{saa2}) yields the low energy potential 
\begin{equation}
V=-G\frac{m E_\gamma}{r}
\label{potphoton}
\end{equation}
which describes the motion of a photon as it had mass equal to its energy in the Newtonian approximation.

Diagram (e) also describes the gravitational redshift.  When the impact parameter vanishes, a photon starting at $-\infty$ approaches the mass target
and experiences the potential (\ref{potphoton}) which accelerates the photon and blueshifts its frequency. 
Thus the energy of the photon at point $r_1$ reads
\begin{equation}
\hbar\omega_1=E_\infty +\frac{GM\hbar\omega_1}{r_1},
\end{equation}
and when $\omega_1>>\omega_2-\omega_1$
\begin{equation}
\frac{\omega_2-\omega_1}{\omega_1} =(\frac{GM }{r_2}-\frac{GM }{r_1})+\mathcal{O}(G^2),
\end{equation}
as required.

Diagram (e) has the same value as its counterpart diagram in General Relativity obtained following the methods of \cite{don} which produces the 
exact same potential.  However, renormalization aside, it does so in the frame of reference where the photon scatters off the Fermion,
and again confirms that we can model photons as particles with mass $E_\gamma$ and (\ref{potphoton}).
Like the gravitational redshift, the Shapiro delay also stems from diagram (e) and it is just the length of the path traveled by the photon when
bending about a mass particle and after reflection from some other body retraces its path back to the source. 

This feature transpires 
the quantum limit were diagrams in General Relativity match the diagrams for QGD to $\mathcal{O}(\frac{v^2}{c^2})$.
In the classical limit, the trajectories of General Relativity on a Schwarzschild background also follow from consideration of a particle moving about a potential. 
Therefore, since we have shown that both theories have the same potential to $\mathcal{O}(\frac{v^2}{c^2})$, any classical or quantum test that General Relativity 
passes will also be passed by QGD to that same order.

The QGD Lagrangian (\ref{lagtodo}) has 4 free parameters $e$, $g$, $g'$ and $\alpha$.   Equations (\ref{eq-a}), (\ref{tree-b}), (\ref{eq-cd}) and (\ref{tree-light}), repeated here:
\begin{eqnarray}
e_{exp}&=&\frac{e}{\beta}, \\
g^{'}&=&(\pi G)^{1/2}  m_{exp},\\
g&=&\frac{128}{27}a \sqrt{\pi G}m,\\
\frac{16}{27}\frac{40\alpha^2}{1+6\alpha^2} &=&\frac{1}{a},
\end{eqnarray}
fix these parameters to experimentally observed quantities from QED, and General Relativity: the electric charge, the Newtonian mass and gravitational constant,
the precession of a mass particle about another mass particle 
and the gravitational lensing of light by mass particles. Therefore, QGD
suffices  to describe the gravitational effects observed in nature without resorting to a spin-2 particle and using instead a quantum connection
defined over an SO(4) fiber bundle.  

The dimensionality of the connecton is the same as that of the U(1) gauge field.  Therefore, the terms in the Lagrangians (\ref{lag12}), (\ref{lag13})
and (\ref{lag2})  involve products of fields and their derivatives with total dimensionality four or less.  
Then all scattering amplitudes considered in this section renormalize by finite amounts\cite{wein2}. 

The matching of the above 4 differential cross sections with those of General Relativity coupled to matter and light are more than simple tests.  By matching the 
differential cross sections we have, at least for two particle interactions, shown that the equations of motion for matter and  light coupled
to General Relativity exactly match those for QGD to $\mathcal{O}(\frac{v^2}{c^2})$. This is manifested by the fact that the Schroedinger equation 
for both theories are the same.   Therefore, any test that follows from General Relativity coupled to matter and light will be reproduced by QGD to order
$\mathcal{O}(\frac{v^2}{C^2})$. While it may very well be the case that General Relativity and QGD diverge beyond $\mathcal{O}(\frac{v^2}{c^2})$,  the reader
is reminded that QGD needs to replicate nature and not General Relativity. We know that because QGD and General Relativity are indistinguishable 
to $\mathcal{O}(\frac{v^2}{c^2})$, it replicates nature to that order.  But General Relativity, safe for \cite{abbott}, has not been experimentally
proven beyond that order, and therefore it may very 
well be that if QGD and General Relativity diverge beyond $\mathcal{O}(\frac{v^2}{c^2})$, it will be up to nature to judge which most closely resembles it, always
taking into account that General Relativity is at best and effective theory because it is not renormalizable like QGD.

It should be further noted that while the differential cross sections pertain two particle interactions and omit single particle and three-, 
four-particle interactions, these do not impose further constraints.  The single particle diagrams involve loops which General Relativity cannot handle 
due to lack of renormalization.  Furthermore, Three and four particle interactions are constrained by two particle scattering; see \cite{don} for 
the diagrams used in General Relativity.  Therefore, by considering
only the equivalency of two particle scattering processes, we fix the equations of motions for all other scattering processes in both theories.

\section{The Standard Model}\label{sm22}

Lack of evidence supporting the existence of a spin-2 particle along with the existence of a mapping
between an SO(4) fiber bundle to an SO(1,3) tangent bundle motivate the formulation
of a gravitational theory based on the SO(4) connecton instead of one based on the graviton.
Without a graviton field and with a constant diagonal metric void of any dynamics,  
QGD again incorporates all the necessary 
gravitational interactions to the Standard Model using instead  $N\times M$ SO(4) gauge fields $\omega_{\mu ab}^{(i,m)}$. 
The inclusion of the fields $\omega_{\mu ab}^{(i,m)}$
suffices to obtain all the experimentally verified differential cross sections, and those expected from General Relativity.
  
The experimental evidence for General Relativity exists only for the matter and U(1) gauge sector.  At present, no experimental evidence exists 
supporting that the 
SU(2), SU(3) and the Higgs sectors of the Standard Model  couple to gravity; however, here we show that QGD can incorporate such couplings to
satisfy the intuitive expectation of General Relativity.

The Lagrangian
\begin{eqnarray}
L&=&L_{gauge}+L_f\label{lagtodoSW}+L_{H}\\
L_{gauge}&=&\frac{1}{4} B_{\mu\nu}B^{\mu\nu}+\frac{1}{4}W_{\mu\nu}^aW^{a\mu\nu}+\frac{1}{4}G_{\mu\nu}^aG^{a\mu\nu}+\frac{1}{4}\sum_{i,m=1}^{m=M} \Omega^{(i,m)}_{\mu\nu ab}\Omega^{(i,m)\mu\nu ab}\nonumber\\
&& \label{lagBSW}\\
L_f&=& \frac{i}{2NM}\sum_{jim}\bar{\psi}^{(j)}\{\gamma^\mu, D^{(i,m)}_\mu\}\psi^{(j)}+
\sum_{jim}\frac{g_{jim}}{g_{im}}\bar{\psi}^{(j)}[\gamma^\mu, D^{(i,m)}_\mu]\psi^{(j)}\label{lagFSW}\\
L_{H}&=&\frac{1}{2\beta_{Higgs}^2}D_\mu \phi\cdot D^\mu \phi -V(\phi\cdot \phi)\label{lagHSW}
\end{eqnarray}
remains invariant under SO(4) transformations as well as $U(1)\times SU(2)\times SU(3)$ transformations.
There are $N\times M$ covariant derivatives $D^{(i,m)}_\mu$ in (\ref{lagFSW}).  This new artifice
allows us to introduce enough parameters to fit all the gravitational scattering amplitudes.

The U(1) gauge field $B_\mu$, the SU(2) gauge field $W^a_\mu$ and the SU(3) gauge field $G^a_\mu$ along with the SO(4) gauge fields $\omega_\mu^{(i,m)ab}$
define the strength fields in (\ref{lagBSW})
\begin{eqnarray}
B_{\mu\nu}&=&\frac{1}{\beta_{U(1)}} [\partial_\mu  B_\nu-\partial_\nu  B_\mu],\\
W_{\mu\nu}^a&=&\frac{1}{\beta_{SU(2)}}\bigg[\partial_\mu  W^a_\nu-\partial_\nu  W^a_\mu
-\frac{g_{SU(2)}}{\beta_{SU(2)}}C^{abc}W_\mu^b W_\nu^c\bigg], \\
G_{\mu\nu}^a&=&\frac{1}{\beta_{SU(3)}} \bigg[\partial_\mu  G^a_\nu-\partial_\nu  G^a_\mu
-\frac{g_{SU(3)}}{\beta_{SU(3)}} C^{abc}G_\mu^b G_\nu^c\bigg], \\
\Omega_{\mu\nu}^{(i,m)ab}&=&(\partial_\mu\omega_\nu^{(i,m)ab}+\frac{\alpha_{im} }{\beta_{im}}\epsilon_\nu^{ab\rho}\partial_\mu A^m_\rho)-
(\partial_\nu\omega_\mu^{(i,m)ab}+\frac{\alpha_{im} }{\beta_{im}}\epsilon_\mu^{ab\rho}\partial_\nu A^m_\rho)\nonumber\\
& &-g_{i\,m} C^{ab}_{\;\;\;cdef}(\omega_\mu^{(i,m)cd}+\frac{\alpha_{im} }{\beta_{im}}\epsilon_\mu^{cd\rho} A^m_\rho)(\omega_\nu^{(i,m)ef}+
\frac{\alpha_{im} }{\beta_{im}}\epsilon_\nu^{ef\sigma} A^m_\sigma)\nonumber\\
&&
\end{eqnarray}
We define each $\tilde\omega_{\mu ab}^{(i,m)}$ as a copy of a textbook SO(4) gauge field necessarily having antisymmetry in indices $a$ and $b$, then,
\begin{equation}
\omega_{\mu ab}^{(i,m)}=\tilde{\omega}_{\mu ab}^{(i,m)}+e^{c}_{\mu} e^{\nu}_{a} \tilde{\omega}_{\nu cb}^{(i,m)},\label{prop2}
\end{equation}
so that
\begin{equation}
\omega_{\mu ab}^{(i,m)}e^{a}_\nu e^{b}_\rho\epsilon^{\mu \nu\rho\sigma}=0.\label{prop2a}
\end{equation}
Since 
\begin{equation}
A^m_\mu=
(B_\mu,W_\mu^1 ,W_\mu^2 ,W_\mu^3,G^1_\mu,G^2_\mu,G^3_\mu,G^4_\mu,G^5_\mu,G^6_\mu,G^7_\mu,G^8_\mu,\phi_\mu),
\end{equation}
where $\phi_\mu=e_\mu^a\phi_a$ and $\phi_a,\; a=1,...,4$ describes the Higgs in the fundamental of SO(4),
the constant couplings $\alpha_{im}$ must take the form  
\begin{equation}
\alpha_{im}=\left\{
\begin{array}{ll} 
\alpha_{iU(1)}\;& m=1\\
\alpha_{iSU(2)}\;& m=2,...,4\\
\alpha_{iSU(3)}\;& m=5,...,12\\
\alpha_{iHiggs}\;& m=13,
\end{array} 
\right\}
\end{equation}
so that 
\begin{equation}
\beta_{im}=\left\{
\begin{array}{lll} 
\beta_{iU(1)}&=\sqrt{N(1+6\alpha_{iU(1)}^2)}\;& m=1\\
\beta_{iSU(2)}&=\sqrt{N(1+6\alpha_{iSU(2)}^2})\;& m=2,...,4\\
\beta_{iSU(3)}&=\sqrt{N(1+6\alpha_{iSU(3)}^2)}\;& m=5,...,12\\
\beta_{iHiggs}&=\sqrt{N(1+6\alpha_{iHiggs}^2)}\;& m=13.
\end{array} 
\right\}
\end{equation}
and 
\begin{equation}
\frac{1}{\beta_{U(1)}^2}=\sum_i \frac{1}{\beta_{iU(1)}^2},\;\;
\frac{1}{\beta_{SU(2)}^2}=\sum_i \frac{1}{\beta_{iSU(2)}^2},\;\;
\frac{1}{\beta_{SU(3)}^2}=\sum_i \frac{1}{\beta_{iSU(3)}^2}.
\end{equation}

When representing a quark doublet, the spinors $\psi^{(j)}$, 
\begin{equation}
\left(
\begin{array}{c}
u_L\\
d_L
\end{array}
\right)
\end{equation}
transform in the irreducible representation of $U(1)\times SU(2)\times SU(3)$ while when 
representing a lepton doublet, the spinors $\psi^{(j)}$, 
\begin{equation}
\left(
\begin{array}{c}
\nu_L\\
e_L
\end{array}
\right)
\end{equation}
transform in the irreducible representation of $U(1)\times SU(2)$.
Then,   Each of the covariant derivatives $D^{(i,m)}$,
\begin{equation}
D^{(i,m)}_\mu=\bigg[\partial_\mu + \frac{ig_{U(1)}}{\beta_{U(1)}} B_\mu + \frac{ig_{SU(2)}}{\beta_{SU(2)}} W_\mu^a\tau^a+\nonumber\\   
  \frac{ig_{SU(3)}}{\beta_{SU(3)}}  G^a_\mu T^a +i \hat g_{im}   \Omega^{(i,m)}_{\mu ab} T^{ab}\bigg],\label{cdSW}
\end{equation}
acting on a quark doublet $\psi^{(j)}$ transforms covariantly under $SO(4)\times U(1)\times SU(2)\times SU(3)$.
The U(1) transformation generated by $\Gamma=e^{i\lambda}$ acts on the following fields
\begin{eqnarray}
\psi'^{(j)}&=&\Gamma\psi'^{(j)}\\
B'_\mu&=&B_\mu-\frac{\beta_{U(1)}}{g_{U(1)}} \partial_\mu\lambda\\
\omega_\mu^{(i,m)'ab}&=& \omega_\mu^{(i,m)ab}+\frac{\alpha_{U(1)}}{g_{U(1)}}\epsilon_\mu^{ab\nu}\partial_\nu\lambda,\label{nov43}
\end{eqnarray}
and leaves (\ref{lagtodoSW}) invariant while each $D^{(i,m)}_\mu\psi^{(j)}$ transforms covariantly.
Since
\begin{equation}
[T^a,T^{ab}]=[\tau^a,T^{ab}]=[T^a,\tau^a]=0,
\end{equation}
SU(3) transformations generated by $\Gamma=e^{i\Lambda^aT^a}$  act on the following fields 
\begin{eqnarray}
\psi'^{(j)}&=&\Gamma\psi'^{(j)}\\
G_\mu^{'a}&=&G_\mu^{a}-\frac{1}{g_{SU(3)}}\partial_\mu\Lambda^a + C^a_{\;bc}\Lambda^b G_\mu^c,
\end{eqnarray}
where $C^a_{\;bc}$ represents the SU(3) structure constant,
to leave (\ref{lagtodoSW}) invariant while each $D^{(i,m)}_\mu\psi^{(j)}$ transforms covariantly.
Similarly, SU(2) transformations generated by $\Gamma=e^{i\Lambda^a\tau^a}$  act on the following fields 
\begin{eqnarray}
\psi'^{(j)}&=&\Gamma\psi'^{(j)}\\
W_\mu^{'a}&=&G_\mu^{a}-\frac{1}{g_{SU(2)}}\partial_\mu\Lambda^a + C^a_{\;bc}\Lambda^b W_\mu^c,
\end{eqnarray}
where $C^a_{\;bc}$ represents the SU(2) structure constant,
and also leave (\ref{lagtodoSW}) invariant while each $D^{(i,m)}_\mu\psi^{(j)}$ transforms covariantly.

SO(4) transformations generated by $\Gamma=e^{i\Lambda_{ab}T^{ab}}$ act on the following fields
\begin{eqnarray}
\psi'&=&\hat\Gamma\psi\\
\omega'^{(i,m)ab}_\mu&=&\omega^{(i,m)ab}_\mu-\frac{1}{g}\partial_\mu\Lambda^{ab}   + C^{ab}_{\;\;\;cdef}\Lambda^{cd}(
\omega_\mu^{(i,m)ef}+\frac{\alpha}{\beta}\epsilon_\mu^{ef\nu}A^m_\nu),
\end{eqnarray}
and also leave (\ref{lagtodoSW}) invariant while each $D^{(i,m)}_\mu\psi^{(j)}$ transforms covariantly.
Under $SO(4)\times U(1)\times SU(2)$ transformations, (\ref{cdSW}) acting on a lepton doublet $\psi^{(j)}$ transforms covariantly 
while (\ref{lagtodoSW}) remains invariant.

The bosonic component of the Lagrangian, (\ref{lagBSW}), 
has quadratic terms\footnote{We omit terms with $A^{13}_\mu$ which are included in (\ref{nov6}) instead.}
\begin{equation}
L^{(2)}_{gauge}=
\frac{1}{2}\partial_\mu B_\nu\partial^\mu B^\nu+
\frac{1}{2}\partial_\mu W^a_\nu\partial^\mu W^{a\nu}+
\frac{1}{2}\partial_\mu G^a_\nu\partial^\mu G^{a\nu}+
\frac{1}{2}\sum_{i,m=1}^{m=M}\partial_\mu \omega^{(i,m)}_\nu\partial^\mu \omega^{(i,m)\nu}.
\label{nov4}
\end{equation}
The dimensionality of the connecton is the same as that of the U(1) gauge field.  Therefore, the terms in the Lagrangian
(\ref{nov4})  involve products of fields and their derivatives with total dimensionality four.  

The pure Standard Model SU(2) and SU(3) cubic and quartic bosonic interactions 
 present in (\ref{lagBSW}) do not mix with the SO(4) gauge fields and do not merit further treatment here.
For the same reasons presented for (\ref{lag13}),
the relevant cubic terms in Lagrangian (\ref{lagBSW}) include
\begin{eqnarray}
L_{gauge}^{(3)}&=&  \sum_{i,m=1}^{m=12} 2g_{i\,m} 
(\partial_\mu\omega_\nu^{(i,m)ab}-\partial_\nu\omega_\mu^{(i,m)ab}) \omega^{(i,m)\mu }_{c a}\; \omega^{(i,m)\nu c }_{\;\;\;b}\nonumber\\
& &
-4g_{i\,m}\frac{\alpha^2_{im} }{\beta^2_{im}} ( A^m_\mu \partial_\rho A^m_\nu e^\nu_a e^\rho_b \omega^{(i,m)\mu a b }+ 
\partial_\nu A^m_\mu A^m_\rho e^\mu_a e^\rho_b\omega^{(i,m)\nu a b}).\nonumber\\
& & 
\label{nov5}
\end{eqnarray}
As in (\ref{lag13}), the terms in (\ref{nov5}) take the form $\omega\omega\partial \omega$ and  $\partial A A \omega$.
In the low energy limit the gauge Boson momentum, $r$, and the connecton momentum, $k$, satisfy $r>>k$.

The terms in the Lagrangian
(\ref{nov5}) also  involve products of fields and their derivatives with total dimensionality four.  
Therefore any diagrams constructed from the terms in (\ref{nov5}) will renormalize by finite amounts\cite{wein2}.

After adding the contributions from (\ref{lagBSW}) dependent on $A^{13}$ to (\ref{lagHSW})
we obtain the Standard Model Higgs Lagrangian along with QGD interactions 
\begin{eqnarray}
L_{Higgs}&=&\frac{1}{2}\partial_\mu \phi \partial^\mu \phi-V(|\phi|^2)\nonumber\\
& & -4\sum_i \frac{\alpha_{i,Higgs}^2}{\beta^2_{i,Higgs}} g_{i\,13} ( \phi_\mu \partial_\rho \phi_\nu e^\nu_a e^\rho_b \omega^{(i,13)\mu a b }+
 \partial_\nu \phi_\mu \phi_\rho e^\mu_a e^\rho_b\omega^{(i,13)\nu a b})\nonumber\\
& & \frac{1}{\beta_{Higgs}} \bigg[ i (\phi^+_d (g_{U(1)}B_\mu+g_{SU(2)}W^a_\mu\tau^a)\partial \phi_d +h.c. \nonumber \\
&  &+\phi_d^+(g_{U(1)}B_\mu+g_{SU(2)}W^a_\mu\tau^a) (g_{U(1)}B_\mu+g_{SU(2)}W^a_\mu\tau^a)\phi_d \bigg],
 \label{nov6}
\end{eqnarray}
where $\phi_d$ describes the Higgs in the SU(2) doublet representation instead of the SO(4) vector representation and 
\begin{equation}
\frac{1}{\beta_{Higgs}^2}=\sum_i \frac{1}{\beta_{iHiggs}^2}.
\end{equation}
The dual representation
of the Higgs further strengthens the proposal of the SO(4) connecton as the carrier of quantum gravitational interactions.

The terms in the Lagrangian
(\ref{nov6})  again involve products of fields and their derivatives with total dimensionality four.  Thus, 
any corrections involving only these terms will renormalize by finite amounts\cite{wein2}.

When $\psi^{(j)}$ represents a quark doublet, the Lagrangian (\ref{lagFSW}) simplifies to
\begin{eqnarray}
L_{f}&=&i
\bar\psi^{(j)}
\Bigg[\slashed \partial+ \frac{i}{\beta_{U(1)}}
\left( \begin{array}{c}
e_{(u)}\\
e_{(d)}
\end{array} \right)
B_\mu \gamma^\mu
+ i \frac{g_2}{\beta_{SU(2)}}W^a_\mu \tau^a\gamma^\mu
+ i \frac{g_3}{\beta_{SU(3)}} G^a_\mu T^a \gamma^\mu\nonumber\\
& & +\sum_{i,m} 
\left( \begin{array}{c}
g_{(u)im}\\
g_{(d)im}
\end{array} \right)
\omega_{\;\;\;\;\mu}^{(i,m)\mu a}\gamma_a 
   + 3i \sum_{i,m} g_i \frac{ \alpha_{im} }{\beta_{im}} A^m_\mu\gamma^5\gamma^\mu \Bigg] 
\psi^{(j)}
,
\label{nov8}
\end{eqnarray}
where $e_{(u)}$ and $e_{(d)}$ are the electric charges of each respective quark and 
\begin{equation}
g_1=\frac{g_{U(1)}}{\beta_{U(1)}},\;
g_2=\frac{g_{SU(2)}}{\beta_{SU(2)}},\;
g_3=\frac{g_{SU(3)}}{\beta_{SU(3)}},
\end{equation}
\begin{equation}
g_{(q)im}=g_{jim}\in \mathbb{C}.
\end{equation}

Similarly, 
when $\psi^{(j)}$ represents a lepton doublet, the Lagrangian (\ref{lagFSW}) simplifies to
\begin{eqnarray}
L_{f}&=&i
\bar\psi^{(j)}
\Bigg[\slashed \partial+  \frac{i}{\beta_{U(1)}}
\left( \begin{array}{c}
e_{(\nu)}\\
e_{(e)}
\end{array} \right)
B_\mu \gamma^\mu
+ i \frac{g_2}{\beta_{SU(2)}}W^a_\mu \tau^a\gamma^\mu
+ i \frac{g_3}{\beta_{SU(3)}} G^a_\mu T^a \gamma^\mu\nonumber\\
& & +\sum_{i,m} 
\left( \begin{array}{c}
g_{(\nu)im}\\
g_{(e)im}
\end{array}
\right)
\omega_{\;\;\;\;\mu}^{(i,m)\mu a}\gamma_a 
   + 3i \sum_{i,m} g_i \frac{ \alpha_{im} }{\beta_{im}} A^m_\mu\gamma^5\gamma^\mu \Bigg] 
\psi^{(j)}
,
\label{nov8b}
\end{eqnarray}
where $e_{(\nu)}$ and $e_{(e)}$ are the electric charges of each respective lepton.

In addition to the transformation of the  terms in the pure Standard Model,
the electroweak rotation which produces the U(1) gauge field $A_\mu$ only transforms
the last term in (\ref{nov8}) and (\ref{nov8b})
\begin{eqnarray}
3i \sum_{i,m} g_i \frac{ \alpha_{im} }{\beta_{im}} A^m_\mu\gamma^5\gamma^\mu&=&
3i \sum_{i} g_i (\frac{ \alpha_{i1} }{\beta_{i1}} cos(\theta_w) +\frac{ \alpha_{i4} }{\beta_{i4}} sin(\theta_w))   A_\mu\gamma^5\gamma^\mu\nonumber\\
 & &3i \sum_{i} g_i (-\frac{ \alpha_{i1} }{\beta_{i1}} sin(\theta_w) +\frac{ \alpha_{i4} }{\beta_{i4}} cos(\theta_w))   Z_\mu\gamma^5\gamma^\mu\nonumber\\
& &+ 3i \sum_{i,m\ne 1,4} g_i \frac{ \alpha_{im} }{\beta_{im}} A^m_\mu\gamma^5\gamma^\mu,
\end{eqnarray}
where $\theta_w$ describes the weak mixing angle and takes the same values as in the pure Standard Model. Thus, for a quark doublet
\begin{eqnarray}
L_{f}&=&L_{f,sm}+
\bar\psi^{(j)}
\Bigg[
\sum_{i,m} 
\left( \begin{array}{c}
g_{(u)im}\\
g_{(d)im}
\end{array}
\right)
\omega_{\;\;\;\;\mu}^{(i,m)\mu a}\gamma_a \nonumber\\
& & 3i \sum_{i} g_i (\frac{ \alpha_{i1} }{\beta_{i1}} cos(\theta_w) +\frac{ \alpha_{i4} }{\beta_{i4}} sin(\theta_w))   A_\mu\gamma^5\gamma^\mu\nonumber\\
& &  +3i \sum_{i} g_i (-\frac{ \alpha_{i1} }{\beta_{i1}} sin(\theta_w) +\frac{ \alpha_{i4} }{\beta_{i4}} cos(\theta_w))   Z_\mu\gamma^5\gamma^\mu
+ 3i \sum_{i,m\ne 1,4} g_i \frac{ \alpha_{im} }{\beta_{im}} A^m_\mu\gamma^5\gamma^\mu,
\Bigg]   
\psi^{(j)}
,\nonumber\\
\label{nov8-2}
\end{eqnarray}
with $L_{f,sm}$ representing the fermionic sector of the pure Standard Model after the rotation between $B_\mu$ and $W^3_\mu$. A similar expression
follows for the lepton doublet.

The dimensionality of the connecton is the same as that of the U(1) gauge field.  Therefore, the terms in the Lagrangian
(\ref{nov8-2})  involve products of fields and their derivatives with total dimensionality of at most four.  
Renormalization then follows\cite{wein2}. 
More importantly, the whole Lagrangian, the sum of (\ref{nov4}), (\ref{nov5}), (\ref{nov6})
and (\ref{nov8-2}), involve products of fields and their derivatives also with total dimensionality at most four
and therefore any corrections involving their terms renormalize by finite amounts\cite{wein2}. 

The electroweak rotation transforms $L_{gauge}^{(2)}$ into 
\begin{eqnarray}
L^{(2)}_{gauge}&=&
\frac{1}{2}\partial_\mu A_\nu\partial^\mu A^\nu+
\frac{1}{2}\partial_\mu W^1_\nu\partial^\mu W^{1\nu}+
\frac{1}{2}\partial_\mu W^2_\nu\partial^\mu W^{2\nu}+
\frac{1}{2}\partial_\mu Z_\nu\partial^\mu Z^\nu+\nonumber\\
& &+\frac{1}{2}\partial_\mu G^a_\nu\partial^\mu G^{a\nu}+
\frac{1}{2}\sum_{i,m=1}^{m=M}\partial_\mu \omega^{(i,m)}_\nu\partial^\mu \omega^{(i,m)\nu},
\label{nov47}
\end{eqnarray}
and it transforms $L_{gauge}^{(3)}$ into
\begin{eqnarray}
L_{gauge}^{(3)}&=&  \sum_{i,m\ne 1,4}^{m=12} 2g_{i\,m}
(\partial_\mu\omega_\nu^{(i,m)ab}-\partial_\nu\omega_\mu^{(i,m)ab}) \omega^{(i,m)\mu }_{c a}\; \omega^{(i,m)\nu c }_{\;\;\;b}\nonumber\\
& &
-4g_{i\,m}\frac{\alpha^2_{im} }{\beta^2_{im}} ( A^m_\mu \partial_\rho A^m_\nu e^\nu_a e^\rho_b \omega^{(i,m)\mu a b }+
\partial_\nu A^m_\mu A^m_\rho e^\mu_a e^\rho_b\omega^{(i,m)\nu a b}).\nonumber\\
& &+ 2 \sum_{i} g_{i\,0}
(\partial_\mu\omega_\nu^{(i,0)ab}-\partial_\nu\omega_\mu^{(i,0)ab}) \omega^{(i,0)\mu }_{c a}\; \omega^{(i,0)\nu c }_{\;\;\;b}\nonumber\\
& &-4\sum_{i}g_{i\,0}\frac{\alpha^2_{i0} }{\beta^2_{i0}} ( A_\mu \partial_\rho A_\nu e^\nu_a e^\rho_b \omega^{(i,0)\mu a b }+ 
\partial_\nu A_\mu A_\rho e^\mu_a e^\rho_b\omega^{(i,0)\nu a b})  \nonumber\\
& &-4\sum_{i}g_{i\,0}\frac{\alpha^2_{i0} }{\beta^2_{i0}} ( Z_\mu \partial_\rho Z_\nu e^\nu_a e^\rho_b \omega^{(i,0)\mu a b }+ 
\partial_\nu Z_\mu Z_\rho e^\mu_a e^\rho_b\omega^{(i,0)\nu a b})  \nonumber\\
& & + terms\; mixing\; Z_\mu\; and \; A_\mu,
\label{nov11}
\end{eqnarray}
after choosing $g_{i\,1}=g_{i\,4}=\sqrt{2} g_{i\,0}$, $\alpha_{i1}=\alpha_{i4}=\alpha_{i0}$  and 
$ \omega^{(i,1)\mu a b }= \omega^{(i,4)\mu a b }= \frac{1}{\sqrt{2}}\omega^{(i,0)\mu a b }$.  

At this point we can  apply the Higgs mechanism
by selecting the usual VEV for the Higgs field which will produce the same effects that take place in  the Standard Model.
The rotation between the fields does not affect the dimensionality of the fields.
Therefore, the rotated terms in the Lagrangian involve products of fields and their derivatives with total dimensionality at most four.  
Renormalization then follows\cite{wein2}.

\section{The U(1) Gauge Boson and Matter Sectors of the Standard Model}

The experimental evidence for gravitational interactions only includes observations for the U(1) gauge fields and the fermionic matter fields. 
Thus, 
while the model can include gravitational interactions with the SU(2), SU(3) and Higgs sectors, we limit ourselves
in this section to the study of the U(1) gauge fields and the fermionic matter fields.  This removes all terms
in (\ref{nov4}), (\ref{nov8-2}), and (\ref{nov11})  which carry index $m$ as well and allows us to drop altogether the Higgs sector (\ref{nov6}).
Then, Lagrangian (\ref{nov8-2}) expressed in terms of true Fermions reduces to
\begin{eqnarray}
L_{f}&=& \sum_j^{J}\bar\psi^{(j)}
\Bigg[(i\slashed \partial-m)-  e_jA_\mu \gamma^\mu +i\sum_{i} g_{ji} \omega_{\;\;\;\;\mu}^{(i)\mu a}\gamma_a \nonumber\\
& &- 3 \sum_{i} g_i \frac{ \alpha_{i} }{\beta_{i}} (cos(\theta_w)  sin(\theta_w))   A_\mu\gamma^5\gamma^\mu\nonumber
\Bigg] 
\psi^{(j)}
,\nonumber\\
& &\label{nov12}
\end{eqnarray}
with $j=1,...,J$ running over the number of singlet spinors,
$g_i=g_{i0}$ and $\alpha_i=\alpha_{i0}$.
The Lagrangian  (\ref{nov47}) reduces to
\begin{equation}
L^{(2)}_{gauge}=
\frac{1}{2}\partial_\mu A_\nu\partial^\mu A^\nu+
\frac{1}{2}\sum_{i}\partial_\mu \omega^{(i)}_\nu\partial^\mu \omega^{(i)\nu},
\label{nov13}
\end{equation}
and the Lagrangian (\ref{nov11}) reduces to
\begin{eqnarray}
L_{gauge}^{(3)}&=&  \sum_{i} 2g_{i}
(\partial_\mu\omega_\nu^{(i)ab}-\partial_\nu\omega_\mu^{(i)ab}) \omega^{(i)\mu }_{c a}\; \omega^{(i)\nu c }_{\;\;\;b}\nonumber\\
& &
-4g_{i}\frac{\alpha_i^2 }{\beta_i^2} ( A_\mu \partial_\rho A_\nu e^\nu_a e^\rho_b \omega^{(i)\mu a b }+
\partial_\nu A_\mu A_\rho e^\mu_a e^\rho_b\omega^{(i)\nu a b}).
\label{nov14}
\end{eqnarray}

The dimensionality of the connecton is the same as that of the U(1) gauge field.  Therefore, the terms in the Lagrangian
(\ref{nov12}), (\ref{nov13}) and (\ref{nov14})  involve products of fields and their derivatives with total dimensionality four.  
Renormalization then follows as shown in\cite{wein2}. 

In sections 2 to 4 we considered a single particle, like the electron, and successfully reproduced the necessary post-Newtonian corrections to fit theory to 
experiment. 
Expanding the particle spectrum increases the number of scattering cross sections well beyond 4; 
This leads to a lack of solution when a single connecton describes the gravitational interactions.  
Instead we introduced $N$ connectons to increase the number of couplings from $g'\in\mathbb{I}$ to $g_{ji}\in\mathbb{C}$ and from $g\in\mathbb{R}$ to 
$g_i\in\mathbb{R}$.  
With this increase in the number of connectons the number of parameters in the model also increases sufficiently to fit all these  differential
cross sections.

The Lagrangian densities (\ref{nov12})-(\ref{nov14}) yield the following propagators
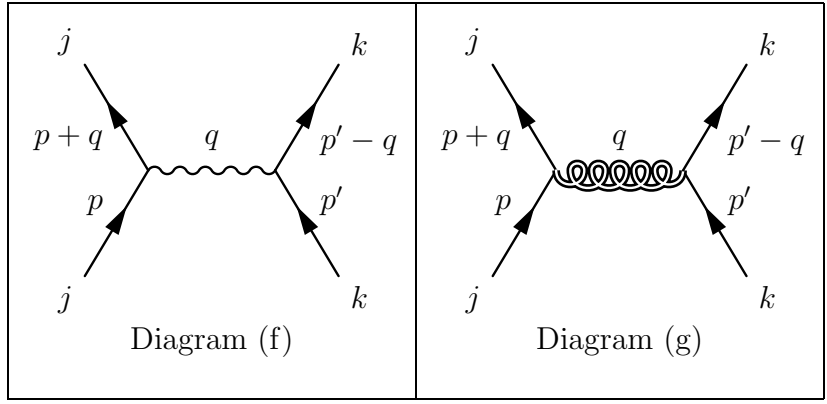
\begin{center}
\begin{tabular}{|p{5cm}|p{4cm}|}
\hline
\begin{center}
\begin{fmffile}{gluonsm}
\begin{fmfgraph*}(70,15)
\fmfleft{in}
\fmfright{out}
\fmflabel{$_\mu^{(i)ab}$}{in}
\fmflabel{$_\nu^{(i)cd}$}{out}
\fmf{dbl_curly,label=$k$,label.dist=10}{in,out}
\end{fmfgraph*}
\end{fmffile}
\end{center}
\begin{center}Propagator (d)\end{center}
 &\vspace{0cm} \Large $-\frac{i g^{\mu\nu}\eta^{ab}\eta^{cd}}{k^2+i\epsilon}$ \vspace{0.5cm} \\ \hline
\begin{center}
\begin{fmffile}{fermionsm}
\begin{fmfgraph*}(70,15)
\fmfleft{in}
\fmfright{out}
\fmflabel{$_j^\alpha$}{in}
\fmflabel{$_j^\beta$}{out}
\fmf{fermion,label=$p$}{in,out}
\end{fmfgraph*}
\end{fmffile}
\end{center}
\begin{center}Propagator (e)\end{center}
& \vspace{0cm}\Large $\frac{i\delta^{\alpha\beta}}{\slashed{p}-m+i\epsilon}$\vspace{0.5cm}  \\ \hline
\begin{center}
\begin{fmffile}{photon}
\begin{fmfgraph*}(70,15)
\fmfleft{in}
\fmflabel{$\mu$}{in}
\fmflabel{$\nu$}{out}
\fmfright{out}
\fmf{photon,label=$r$}{in,out}
\end{fmfgraph*}
\end{fmffile}
\end{center}
\begin{center}Propagator (f)\end{center}
 &\vspace{0cm}\Large $-\frac{ig^{\mu\nu}}{r^2+i\epsilon}$\vspace{0.5cm}\\
\hline
\end{tabular}
\end{center}
For the avoidance of doubt, in the low energy limit 
\begin{equation}
m>>\partial A>>\partial\omega
\end{equation}
such that  $m>>r>>k$.
These same Lagrangian densities generate the following vertices:
\begin{center}
\begin{tabular}{ |p{5.5cm}| p{7.5cm}| }
\hline
\begin{center}
\begin{fmffile}{vertex1sm}
\begin{fmfgraph*}(100,70)
\fmfleft{i1,i2}
\fmfright{o1}
\fmf{fermion,tension=0.5}{i1,v1}
\fmf{fermion,tension=0.5}{v1,i2}
\fmf{dbl_curly,tension=0.5}{v1,o1}
\fmflabel{$_j^\alpha$}{i1}
\fmflabel{$_j^\beta$}{i2}
\fmflabel{$_\mu^{(i)ab}$}{o1}
\end{fmfgraph*}
\end{fmffile}
\end{center}
\begin{center}Vertex (e)\end{center}
&\vspace{1cm}\large $ -g_{ji} \; e^{\mu a} e^{\nu b} \gamma_\nu $ \\ [7ex]\hline
\end{tabular}
\end{center}
\begin{center}
\begin{tabular}{ |p{5.5cm}| p{8.5cm}| }
\hline
\begin{center}
\begin{fmffile}{vertex2sm}
\begin{fmfgraph*}(100,70)
\fmfleft{i1,i2}
\fmfright{o1}
\fmf{fermion,tension=0.5}{i1,v1}
\fmf{fermion,tension=0.5}{v1,i2}
\fmf{photon,tension=0.5}{v1,o1}
\fmflabel{$_j^\alpha$}{i1}
\fmflabel{$_j^\beta$}{i2}
\fmflabel{$_\mu$}{o1}
\end{fmfgraph*}
\end{fmffile}
\end{center}
\begin{center}Vertex (f)\end{center}
&\vspace{1cm} \large $ -\frac{i }{\beta} (e \gamma_\mu+3g_i\alpha_i \gamma_5\gamma_\mu)  $ \vspace{.5cm} \\[7ex] \hline
\begin{center}
\begin{fmffile}{vertex3sm}
\begin{fmfgraph*}(100,70)
\fmfleft{i1,i2}
\fmfright{o1}
\fmf{dbl_curly,tension=1.5,label=$p$,label.dist=-15}{i1,v1}
\fmf{dbl_curly,tension=1.5,label=$q$,label.dist=-15}{v1,i2}
\fmf{dbl_curly,tension=1.5,label=$r$,label.dist=-15}{v1,o1}
\fmflabel{$_\mu^{(i)ab}$}{i1}
\fmflabel{$_\nu^{(i)cd}$}{i2}
\fmflabel{$_\rho^{(i)ed}$}{o1}
\end{fmfgraph*}
\end{fmffile}
\end{center}
\begin{center}Vertex (g)\end{center}
 &
\vspace{1cm}
\begin{tabular}{c}
\small$g_i(h^{af}h^{ce}h^{bd}+h^{ad}h^{ce}h^{bf}-h^{ae}h^{cb}h^{df}+h^{ac}h^{eb}h^{df})\cdot$\\
 $\cdot((p-q)^\rho g^{\mu\nu}+(q-r)^\mu g^{\nu\rho}+(r-p)^\nu g^{\mu\rho})$
\end{tabular}\\[7ex] \hline
\begin{center}
\begin{fmffile}{vertex4sm}
\begin{fmfgraph*}(100,70)
\fmfleft{i1,i2}
\fmfright{o1}
\fmf{photon,tension=1.5,label=$r$}{i1,v1}
\fmf{photon,tension=1.5,label=$r'$}{v1,i2}
\fmf{dbl_curly,tension=1.5,label=$q$,label.dist=-15}{v1,o1}
\fmflabel{$_j^\beta$}{i2}
\fmflabel{$_j^\alpha$}{i1}
\fmflabel{$_\mu^{(i)ab}$}{o1}
\end{fmfgraph*}
\end{fmffile}
\end{center}
\begin{center}Vertex (h)\end{center}
&\vspace{1cm}\large $ g_i\frac{4\alpha_i^2}{\beta_i^2}(g_{\mu\alpha}e^a_\beta r^{'b}+ g_{\mu\beta}e^a_\alpha r^{b} 
+e^a_\alpha e^b_\beta r_\mu   +e^a_\beta e^b_\alpha r'_\mu) $ \\ [7ex]\hline
\end{tabular}
\end{center}

\begin{center}
\begin{tabular}{ |p{5cm}| p{5cm}| }
\hline
\begin{center}
\begin{fmffile}{dia2bis}
\begin{fmfgraph*}(120,80)
\fmfleft{i1,i2}
\fmfright{i3,i4}
\fmf{fermion,tension=0.2,label=$p$,label.side=left}{i1,v1}
\fmf{fermion,tension=0.2,label=$p+q$,label.side=left}{v1,i2}
\fmf{fermion,tension=0.2,label=$p'$,label.side=right}{i3,v2}
\fmf{fermion,tension=0.2,label=$p'-q$,label.side=right}{v2,i4}
\fmf{photon,tension=0.2,label=$q$,label.dist=-15}{v1,v2}
\fmflabel{$j$}{i1}
\fmflabel{$k$}{i3}
\fmflabel{$j$}{i2}
\fmflabel{$k$}{i4}
\end{fmfgraph*}
\end{fmffile}
\end{center}
\begin{center}Diagram (f)\end{center}
&
\begin{center}
\begin{fmffile}{dia1bis}
\begin{fmfgraph*}(120,80)
\fmfleft{i1,i2}
\fmfright{i3,i4}
\fmf{fermion,tension=0.2,label=$p$,label.side=left}{i1,v1}
\fmf{fermion,tension=0.2,label=$p+q$,label.side=left}{v1,i2}
\fmf{fermion,tension=0.2,label=$p'$,label.side=right}{i3,v2}
\fmf{fermion,tension=0.2,label=$p'-q$,label.side=right}{v2,i4}
\fmf{dbl_curly,tension=0.2,label=$q$,label.dist=-15}{v1,v2}
\fmflabel{$j$}{i1}
\fmflabel{$k$}{i3}
\fmflabel{$j$}{i2}
\fmflabel{$k$}{i4}
\end{fmfgraph*}
\end{fmffile}
\end{center}
\begin{center}Diagram (g)\end{center}
\\
\hline
\end{tabular}
\end{center}

\vspace{1cm}

After assuming as before that the axial current vanishes, 
the scattering process in diagram (f)  describing the photon exchange between particle $j$ and $k$ produces a T-matrix element,
\begin{equation}
T_{QED}^{jk} =-4i m_jm_{k}\frac{e_j e_k}{-q^2},\label{tqed}
\end{equation}
which in turn yields a  differential cross section contribution, 
\begin{equation}
\frac{d\sigma_{QED}^{jk}}{d\Omega}= \frac{4 \mu_{jk}^2(e_j e_k/(4\pi))^2}{-q^2},\label{sqed}
\end{equation}
with $\mu_{jk}$ defined as the reduce mass of particles $j$ and $k$. 
On the other hand, the 
Schrodinger equation for a Coulomb potential 
\begin{equation}
V_{C}= \frac{e_{j,exp}e_{k,exp}}{4\pi R},
\end{equation}
with $e_{j, exp}$ and $e_{k, exp}$ describing the particles measured electric charge and $R$ describing 
the distance between the two particles, yields 
the differential cross section 
\begin{equation}
\frac{d\sigma_{C}^{jk}}{d\Omega}=
 \frac{4 \mu_{jk}^2(e_{j,exp} e_{k,exp}/(4\pi))^2}{-q^2}.\label{coulomb-abis}
\end{equation}
Comparing the differential cross section (\ref{sqed}) and (\ref{coulomb-abis}) constrains the parameters $e_j$ and $e_k$ to
\begin{equation}
e_{j}=e_{j,exp}, \;\; e_k=e_{k,exp}.\label{eqpn10}
\end{equation}

Diagram (g), exchanging a connecton between particle $j$ and $k$ produces the T-matrix, 
\begin{equation}
T_{QGD,N}^{jk} =-4i m_jm_{k} \frac{4 \sum^N_i g_{ji}g_{ki}}{-q^2},\label{tqgd}\\
\end{equation}
which in turn produces the differential cross section, 
\begin{equation}
\frac{d\sigma_{QGD,N}^{jk}}{d\Omega}= \frac{4 \mu_{jk}^2\bigg|\sum_i^N g_{ji}g_{ki}\bigg|^2/\pi^2}{q^4}.\label{sqgd}
\end{equation}
On the other side, the differential cross section obtained from the 
Schrodinger equation for a Newton potential, 
\begin{equation}
V_{N}=-G \frac{m_{j,exp}e_{k,exp}}{ R},
\end{equation}
where $m_{j, exp}$ and $m_{k, exp}$  describe the particles measured mass and $R$ the distance between them, produces the differential cross section,
\begin{equation}
\frac{d\sigma_{N}^{jk}}{d\Omega}=
 \frac{4 \mu_{jk}^2(G m_{j,exp} m_{k,exp})^2}{q^4}.\label{newton-abis}
\end{equation}
Comparing the differential cross sections (\ref{sqgd}) and (\ref{newton-abis}) imposes the constraint
\begin{equation}
\bigg|\sum_i^N g_{ji}g_{ki}\bigg|=\pi G m_{j,exp} m_{k,exp}.\label{eqpn11}
\end{equation}

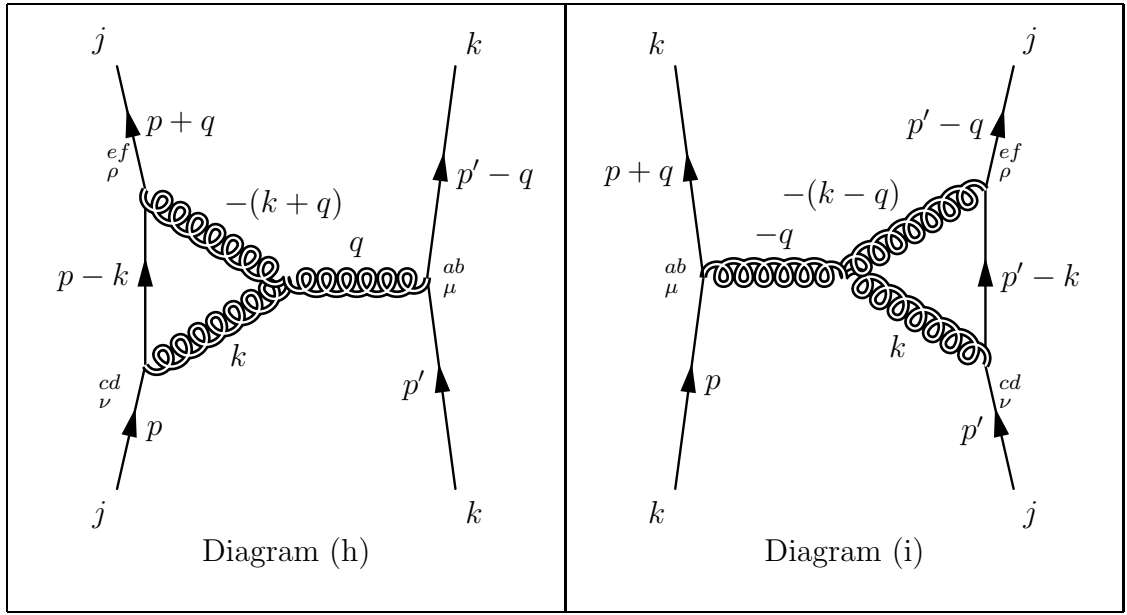
\begin{center}
\begin{tabular}{|p{7cm}|p{7cm}|}
\hline

\begin{center}
\begin{fmffile}{dia3bis}
\begin{fmfgraph*}(160,160)

\fmfleft{i1,i2}
\fmfright{j1,j2}
\fmflabel{$j$}{i1}
\fmflabel{$j$}{i2}
\fmflabel{$k$}{j1}
\fmflabel{$k$}{j2}

\fmf{fermion,label=$p-k$,label.side=left,tension=0.6}{v1,v2}
\fmf{fermion,label=$p+q$,label.side=right}{v2,i2}
\fmf{fermion,label=$p$}{i1,v1}

\fmf{dbl_curly,label=$k$,tension=.2,label.side=right,label.dist=10}{v1,v3}
\fmf{dbl_curly,tension=.2,label=$-(k+q)$,label.side=left}{v2,v3}
\fmf{dbl_curly,label=$q$,tension=.4,label.dist=-15}{v3,v4}
\fmf{fermion,label=$p'$}{j1,v4}
\fmf{fermion,label=$p'-q$}{v4,j2}

\fmflabel{$^{ef}_\rho$}{v2}
\fmflabel{$^{cd}_\nu$ }{v1}
\fmflabel{$^{ab}_\mu$}{v4}

\end{fmfgraph*}
\end{fmffile}
\end{center}
\begin{center}Diagram (h)\end{center}
&
\begin{center}
\begin{fmffile}{dia32bis}
\begin{fmfgraph*}(160,160)

\fmfleft{j1,j2}
\fmfright{i1,i2}
\fmflabel{$j$}{i1}
\fmflabel{$j$}{i2}
\fmflabel{$k$}{j1}
\fmflabel{$k$}{j2}

\fmf{fermion,label=$p'-k$,label.side=right,tension=0.6}{v1,v2}
\fmf{fermion,label=$p'-q$,label.side=left}{v2,i2}
\fmf{fermion,label=$p'$}{i1,v1}

\fmf{dbl_curly,label=$k$,tension=.2,label.side=left}{v1,v3}
\fmf{dbl_curly,tension=.2,label=$-(k-q)$,label.side=right,label.dis=10}{v2,v3}
\fmf{dbl_curly,label=$-q$,tension=.4,label.dist=-20}{v3,v4}
\fmf{fermion,label=$p$}{j1,v4}
\fmf{fermion,label=$p+q$}{v4,j2}
\fmflabel{$^{ef}_\rho$}{v2}
\fmflabel{$^{cd}_\nu$ }{v1}
\fmflabel{$^{ab}_\mu$}{v4}
\end{fmfgraph*}
\end{fmffile}
\end{center}
\begin{center}Diagram (i)\end{center}
\\
\hline
\end{tabular}
\end{center}
\vskip 1.0cm

Diagrams (h) and (i) produce the 1-loop correction to the connecton exchange between particles $j$ and $k$ to the T-matrix,  
\begin{equation}
T_{QGD,E}^{jk}=-4im_jm_k \frac{3}{64} \frac{\sum_i^N g_ig_{ji}g_{ki} ( g_{ji}( m_j+17 m_k)+g_{ki}(17  m_j + m_k))}{m_jm_k\sqrt{-q^2}},
\label{solve2}
\end{equation}
after using the same steps used in section 4, and in turn it produces the differential cross section,
\begin{equation}
\frac{\sigma^{jk}_{QGD,E}}{d\Omega}=
\frac{4\mu^2_{jk}}{|q|^2}
\bigg(\frac{3}{64} \frac{\bigg|\sum_i^N g_ig_{ji}g_{ki}  ( g_{ji}( m_j+17 m_k)+g_{ki}(17  m_j + m_k))\bigg|}{16\pi^2m_jm_k}\bigg)^2.
\label{solve22}
\end{equation}
The Schroedinger equation with the Einstein correction to the gravitational potential between two particles, 
\begin{equation}
V_{E}=- a G^2 \frac{   m_{j,exp}m_{k,exp}(m_{j,exp}+m_{k,exp})}{R^2}, 
\end{equation}
yields the differential scattering cross section 
\begin{equation}
\frac{\sigma_{E}}{d\Omega}=\frac{4\mu_{jk}^2}{|q|^2}  4\pi^2 a^2 G^4 (m_{j,exp}m_{k,exp}(m_{j,exp}+m_{k,exp}))^2.
\label{einstein-cdbis}
\end{equation}
Comparing the differential cross section (\ref{einstein-cdbis}) and (\ref{solve22}) yields
\begin{equation}
\bigg|\sum_i^N g_ig_{ji}g_{ki} ( g_{ji}( m_j+17 m_k)+g_{ki}(17  m_j + m_k))\bigg| =
\frac{16^3\pi^3}{6} a G^2 m^2_{j,exp}m^2_{k,exp}(m_{j,exp}+m_{k,exp}).\label{eqpn12}
\end{equation}

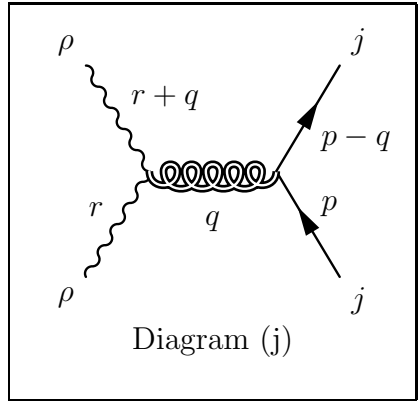
\begin{center}
\begin{tabular}{ |p{5cm}| }
\hline
\begin{center}
\begin{fmffile}{dia6bis}
\begin{fmfgraph*}(120,80)
\fmfleft{i1,i2}
\fmfright{i3,i4}
\fmflabel{$j$}{i3}
\fmflabel{$j$}{i4}
\fmf{photon,tension=0.2,label=$r$,label.side=left}{i1,v1}
\fmf{photon,tension=0.2,label=$r+q$,label.side=right}{v1,i2}
\fmf{dbl_curly,tension=.2,label=$q$,label.dist=15}{v1,v2}
\fmf{fermion,tension=0.2,label=$p$,label.side=right}{i3,v2}
\fmf{fermion,tension=0.2,label=$p-q$,label.side=right}{v2,i4}
\fmflabel{$\rho$}{i1}
\fmflabel{$\rho$}{i2}
\end{fmfgraph*}
\end{fmffile}
\end{center}
\begin{center}Diagram (j)\end{center}
\\
\hline
\end{tabular}
\end{center}

The T-matrix contribution from diagram (j) scattering a mass particle off of a photon simplifies to
\begin{equation}
T_{f_j\,\gamma\to f_j\,\gamma}=2im_j\frac{40 \alpha_i^2}{\beta_i^2} \sum_i g_ig_{ji}  \frac{ E_{\gamma}}{-q^2}, \label{sab0}
\end{equation}
After using the small angle approximation, the differential cross section obtained from this T-matrix equates to 
\begin{equation}
\frac{d\sigma_{f_j\,\gamma\to f_j\,\gamma}}{d\Omega}=\frac{1}{64\pi^2 }\bigg|\sum_i\bigg(\frac{40 \alpha_i^2}{\beta_i^2}\bigg)^2
g_ig_{ji}\bigg|^2 \frac{E_{\gamma}^2}{(r^2 sin^2(\theta/2))^2}.\label{sab1}
\end{equation}
In the low energy limit, equation (\ref{sab1}) describes the Rutherford scattering of a mass $m_j$ projectile off of a
mass $E_{\gamma}$ target.

For a small angle deflection in General Relativity the 
gravitational lensing of a photon by a point particle of mass $m_j$, the differential cross section calculation follows in the same manner as in section 4.3 and resulting in
\begin{equation}
\frac{d\sigma_{L}}{d\Omega} =G^2 E_{\gamma}^2\frac{m_j^2 E_{\gamma}^2}{(r^2sin^2(\theta/2))^2}. 
\label{sab4}
\end{equation}
The exchange $m_j \leftrightarrow E_{\gamma}$
transforms the differential cross section to describe the
scattering of a particle of mass $m_j$ from a photon 
target with energy $E_{\gamma}$ and as a result takes the form
\begin{equation}
\bigg(\frac{d\sigma_{L}}{d\Omega}\bigg)_{m \leftrightarrow E_{\gamma}} =G^2 m_j^2\frac{m_j^2 E_{\gamma}^2}{(r^2sin^2(\theta/2))^2}. 
\label{sab5}
\end{equation}

Equating (\ref{sab1}) with (\ref{sab5}) requires that 
\begin{equation}
\frac{1}{8\pi G m_j^2}\bigg|\sum_i\frac{40 \alpha_i^2}{N(1+6\alpha_i^2)}g_ig_{ji}\bigg|
 =1.
\label{eqpn13}
\end{equation}

Renormalization follows by construction because the structure added to the Standard Model is just $N$ standard Yang-Mills gauge fields with $SO(4)$
symmetry.  Dimensional analysis
of the Standard Model coupled to QGD shows 
that the connectons $\omega^{(i,m)ab}_\mu$ have the same dimension as the U(1), SU(2) and SU(3) gauge fields. 
Therefore, all integrands in the Standard Model coupled to QGD, including those involving the connectons $\omega^{(i,m)}$
have the dimension four or less which  guarantees renormalization of all corrections to physical quantities\cite{wein2}.

If a solution also exists for the system of equations (\ref{eqpn11}), (\ref{eqpn12}) and (\ref{eqpn13}), then the renormalizable Lagrangian
(\ref{nov12}), (\ref{nov13}) and (\ref{nov14}) correctly describes experimental observation without resorting to the graviton and 
instead using $SO(4)$ gauge fields.

If a solution exists to the system of equations 
\begin{eqnarray}
\sum_i^N g_{ji}g_{ki}&=&\pi G m_{j,exp} m_{k,exp},\label{eqqn11}\\
\sum_i^N g_ig_{ji}g_{ki} ( g_{ji}( m_j+17 m_k)+g_{ki}(17  m_j + m_k)) &=&
\frac{16^3\pi^3}{6} a G^2 m^2_{j,exp}m^2_{k,exp}(m_{j,exp}+m_{k,exp}),\nonumber\\
 \label{eqqn12}\\
\frac{1}{8\pi G m_j^2}\sum_i\frac{40 \alpha_i^2}{N(1+6\alpha_i^2)}g_ig_{ji}&=&1,\label{eqqn13}
\end{eqnarray}
then, a solution also exists for the system of equations (\ref{eqpn11}), (\ref{eqpn12}) and (\ref{eqpn13}).

Equation (\ref{eqqn11}) is a stricter for of (\ref{eqpn11}), (\ref{eqqn12}) is a stricter for of (\ref{eqpn12}) and 
(\ref{eqqn13}) is a stricter for of (\ref{eqpn13}). While (\ref{eqpn11}), (\ref{eqpn12}) and (\ref{eqpn13}) are all real, the complex system of 
equations (\ref{eqqn11}), (\ref{eqqn12}) and (\ref{eqqn13}) require their imaginary components to vanish. Then, the solution to
the real components of the system 
of equations (\ref{eqqn11}), (\ref{eqqn12}) and (\ref{eqqn13}) 
also solves the system of equations (\ref{eqpn11}), (\ref{eqpn12}) and (\ref{eqpn13}).

We require the complex system of equations (\ref{eqqn11}) with $\bold{g}_j\in \mathbb{C}^N$, $\bold{g}_j=(g_{j1},...,g_{jN})$ instead
of (\ref{eqpn11}) because solutions to (\ref{eqqn11}) ensure that the $\mathcal{O}(G)$ correction to the $j-$Fermion self energy, 
\begin{equation}
\Sigma^{(G)}_j\propto\sum_i^N g_{ji}^2
\end{equation}
does not have an imaginary component, so that $\Sigma^{(G)}_j$  does not affect unitarity to $\mathcal{O}(G)$.

To show existence of a solution to the system (\ref{eqqn11}), (\ref{eqqn12}) and (\ref{eqqn13}) first note that (\ref{eqqn12}) and  (\ref{eqqn13})
is a complex linear equation system in $g_i\in\mathbb{R}$ and then assume that $\alpha_i=1$. Our model contains 12 Fermions with $j=1,...,J=12$, 
so that the linear system (\ref{eqqn12}) has  $(12\times 13)/2$ real equations and $(12\times 13)/2$ imaginary equations. The linear 
system (\ref{eqqn13}) has 12 real and
12 imaginary equations.  We can also represent the complex linear system (\ref{eqqn12}) and (\ref{eqqn13}) instead 
as a real system of 180 linear equations 
\begin{equation}
\bold{A}\cdot \bold{g}=\bold{b}\label{lsys1}
\end{equation}
where $\bold{g}=(g_1,...,g_N)$ and $\bold A$ represents a square matrix with dimension N=180. 
Then, a solution to (\ref{lsys1}) exists whenever the 
$determinant(\bold{A})\ne 0$ or equivalently when the vector rows in $\bold A$ are not parallel among themselves. 
A simple inspection of the coefficients
of $\bold{A}$ produced by (\ref{eqqn12}) and (\ref{eqqn13}) shows that when the complex vectors $\bold{g}_j$ are not parallel among themselves, the
different vector rows in $\bold{A}$ are not parallel.  A solution to the complex system of equations (\ref{eqqn11}) can be obtained  for
the set $\bold{g}_j\in\mathbb{C}^{180},\;j=1,..,12$ following these steps: 1) assign a real random number between 0 and 1 to each $Re(g_{ji})$
and $Im(g_{ji})$, $j=1,...,12;\;i=1,...,N$, 2) use the Fletcher-Reeves-Polak-Ribiere algorithm to minimize the equation
\begin{eqnarray}
\sum_{jk}^{12}(\sum_i^N Re(g_{ji})Re(g_{ki})-Im(g_{ji})Im(g_{ki})-\pi G m_jm_k)^2+\nonumber\\
	+(\sum_i^N Im(g_{ji})Re(g_{ki})+Im(g_{ki})Re(g_{ji}))^2.\label{lsys2}
\end{eqnarray}
This method produces a minimum of zero for this equation which amounts to solving (\ref{eqqn11}). Note that the first squared term
in (\ref{lsys2}) is the real part of (\ref{eqqn11}) while the second squared term is the imaginary part of (\ref{eqqn11}).
It is not surprising that a minimum of zero exists for (\ref{lsys2}); there are $12\times(180+180)$ variables and only 88 equations.
Furthermore, the vectors produced by this
method are not parallel and explicit calculation of $\bold{A}$ shows that its determinant does not vanish\footnote{The number of couplings can be further reduced
in the limit that $g_{i=1}=0$. Then, we can choose $g_{ji=1}\in \mathbb{I},\; g_{ji\ne1}\in \mathbb{R}$ effectively reducing the number of couplings 
to matter and the number of connectons by half.}.

Therefore, a solution can always be found to the system of 12 Fermions and one U(1) gauge Boson provided there are 180 connectons.   
This does not preclude the existence of solutions whenever $N<180$, specially after abandoning the assumption that $\alpha_i=1$.
 
The reader should be concerned by the number of parameters necessary to fit QGD coupled to the Standard Model when compared with General Relativity.
However, two points should be kept in mind.  First, the intention of this section was to show the existence of a solution rather than to find a solution
with the smallest possible number of couplings. Given the non-linear nature of the system of equations, it is very likely that non-linear solutions
with a significantly smaller number of couplings may very well exist;
for example solutions obtained through the gradient method or simulated annealing. 
Second, QGD should only be compared with theories that successfully incorporate gravity and more
importantly renormalize.  Thus, when compared with heterotic strings, which have 496 gauge fields before even considering compactification, the number of
gauge fields required by QGD is similar. Or, alternatively, like string theory, the gauge fields may represent some compactification of M-theory.  

Nevertheless, we can cast QGD in a more symmetric manner
where each of the 25 particles, whether a Fermion or a gauge Boson, 
couples to 25 different connectons; each particle carries charge with respect to 25 non-Abelian gauge fields;
each particle connects to a different particle through a unique connecton, 
including one which self couples. Then in this construction a picture emerges
where each of the 325 connectons mediates the interaction between two particular particles only (for example between a Higgs and
a $\nu_\tau$). Then, the number, $25\times13+25\times13=650$, of differential cross sections to $\mathcal{O}(\frac{v^2}{c^2})$ 
is smaller than the number of couplings which, when omitting the $\alpha$'s, counts to $25\times25+325=950$: the first term
counts the covariant derivative couplings and the second term counts the Yang Mills couplings.

\section{Bending the Remaining Bosons}
Experimental data evidencing the gravitational interactions does not yet exist for the  $W^{\pm}$, the $Z$ and the Higgs particles or the Gluons. 
However, we expect for them to become available in the future. Nevertheless, QGD can incorporate gravitational interactions for all these 
particles through the second
term in (\ref{nov5}).  We present the propagators for the remainder of the bosonic sector 
\begin{center}
\begin{tabular}{|p{5cm}|p{6cm}|}
\hline
\begin{center}
\begin{fmffile}{gluonB}
\begin{fmfgraph*}(70,15)
\fmfleft{in}
\fmfright{out}
\fmflabel{$m,\mu$}{in}
\fmflabel{${m'},\nu$}{out}
\fmf{gluon,label=$r$,label.dist=10}{in,out}
\end{fmfgraph*}
\end{fmffile}
\end{center}

\begin{center}Propagator (g)\end{center}
 &\vspace{0cm} {\Large $-\frac{i g^{\mu\nu}\delta^{mm'}}{r^2+i\epsilon}$}, $m=5,...,12$ \vspace{0.5cm} \\ \hline
\begin{center}
\begin{fmffile}{WB}
\begin{fmfgraph*}(70,15)
\fmfleft{in}
\fmfright{out}
\fmflabel{$m,\mu$}{in}
\fmflabel{${m'},\nu$}{out}
\fmf{photon,label=$r$}{in,out}
\end{fmfgraph*}
\end{fmffile}
\end{center}
\begin{center}Propagator (h)\end{center}
& \vspace{0cm}{\Large $-\frac{ig^{\mu\nu}\delta^{mm'}}{r^2-M_m^2+i\epsilon}$}, $m=W,Z$\vspace{0.5cm}  \\ \hline
\begin{center}
\begin{fmffile}{HiggsB}
\begin{fmfgraph*}(70,15)
\fmfleft{in}
\fmflabel{$a$}{in}
\fmflabel{$b$}{out}
\fmfright{out}
\fmf{dashes_arrow,tension=2,label=$r$}{in,out}
\end{fmfgraph*}
\end{fmffile}
\end{center}
\begin{center}Propagator (i)\end{center}
 &\vspace{0cm}{\Large $-\frac{i\delta^{ab}}{r^2-M_H^2+i\epsilon}$}\vspace{0.5cm}\\
\hline
\end{tabular}
\end{center}
where the first propagator describes the Gluons, 
the second describes the electroweak vector Bosons $W$ and $Z$,  and  the third describes the Higgs in the SO(4) representation. 

Lagrangian (\ref{nov5}) produces the following vertices: 
\begin{center}
\begin{tabular}{ |p{7cm}| p{7.5cm}| }
\hline
\begin{center}
\begin{fmffile}{vertex5}
\begin{fmfgraph*}(100,70)
\fmfleft{i1,i2}
\fmfright{o1}
\fmf{gluon,tension=1.5,label=$r$}{i1,v1}
\fmf{gluon,tension=1.5,label=$r'$}{v1,i2}
\fmf{dbl_curly,tension=1.5,label=$q$,label.dist=-15}{v1,o1}
\fmflabel{$m,\beta$}{i2}
\fmflabel{$m',\alpha$}{i1}
\fmflabel{$_\mu^{(i,m)ab}$}{o1}
\end{fmfgraph*}
\end{fmffile}
\end{center}
\begin{center}Vertex (i)\end{center}
&\vspace{1cm}\large $\delta^{mm'} g_{im}\frac{4\alpha_{i,SU(3)}^2}{\beta_{i,SU(3)}^2}(g_{\mu\alpha}e^a_\beta r^{'b}+ g_{\mu\beta}e^a_\alpha r^{b} 
+e^a_\alpha e^b_\beta r_\mu   +e^a_\beta e^b_\alpha r'_\mu) $ \\ [7ex]\hline
\begin{center}
\begin{fmffile}{vertex6}
\begin{fmfgraph*}(100,70)
\fmfleft{i1,i2}
\fmfright{o1}
\fmf{photon,tension=1.5,label=$r$}{i1,v1}
\fmf{photon,tension=1.5,label=$r'$}{v1,i2}
\fmf{dbl_curly,tension=1.5,label=$q$,label.dist=-15}{v1,o1}
\fmflabel{$m,\beta$}{i2}
\fmflabel{$m',\alpha$}{i1}
\fmflabel{$_\mu^{(i,m)ab}$}{o1}
\end{fmfgraph*}
\end{fmffile}
\end{center}
\begin{center}Vertex (j)\end{center}
&\vspace{1cm}\large {$ \delta^{mm'}g_{im}\frac{4\alpha_{iSU(2)}^2}{\beta_{iSU(2)}^2}(g_{\mu\alpha}e^a_\beta r^{'b}+ g_{\mu\beta}e^a_\alpha r^{b}
+e^a_\alpha e^b_\beta r_\mu   +e^a_\beta e^b_\alpha r'_\mu)$},$\; m=W,Z $ \\ [7ex]\hline
\begin{center}
\begin{fmffile}{vertex7}
\begin{fmfgraph*}(100,70)
\fmfleft{i1,i2}
\fmfright{o1}
\fmf{dashes_arrow,tension=1.5,label=$r$}{i1,v1}
\fmf{dashes_arrow,tension=1.5,label=$r'$}{v1,i2}
\fmf{dbl_curly,tension=1.5,label=$q$,label.dist=-15}{v1,o1}
\fmflabel{$d$}{i2}
\fmflabel{$c$}{i1}
\fmflabel{$_\mu^{(i,H)ab}$}{o1}
\end{fmfgraph*}
\end{fmffile}
\end{center}
\begin{center}Vertex (k)\end{center}
&\vspace{1cm}\large $ g_{iH}\frac{4\alpha_{iH}^2}{\beta_{iH}^2}(g_{\mu c}e^a_d r^{'b}+ g_{\mu d}e^a_c r^{b}
+e^a_c e^b_d r_\mu   +e^a_d e^b_c r'_\mu) $ \\ [7ex]\hline
\end{tabular}
\end{center}
In general we expect that Gluon couplings satisfy $g_{im}=g_{i,SU(3)},\,m=5,...,12$; 
but, due to the mass difference between the W and Z particles, we expect  $g_{iW}\ne g_{iZ}$ and $g_{jiW}\ne g_{jiZ}$. 

These additional vertices produce the following tree-level and $\mathcal{O}(G)$ diagrams: 
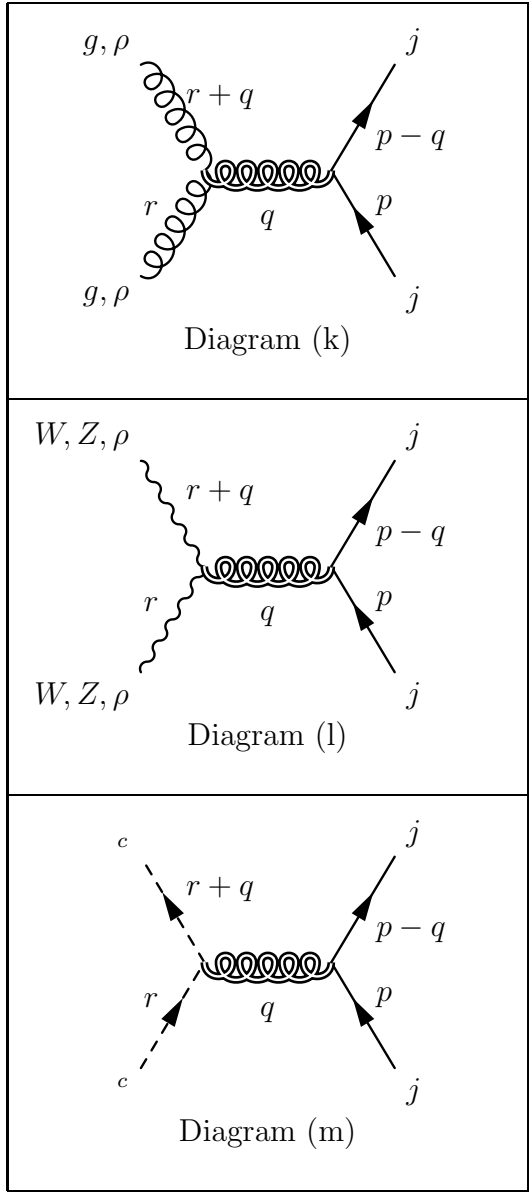
\begin{center}
\begin{tabular}{ |p{6.5cm}|p{6.5cm}| }
\hline
\begin{center}
\begin{fmffile}{dia6G}
\begin{fmfgraph*}(120,80)
\fmfleft{i1,i2}
\fmfright{i3,i4}
\fmflabel{$j$}{i3}
\fmflabel{$j$}{i4}
\fmf{gluon,tension=0.2,label=$r$,label.side=left}{i1,v1}
\fmf{gluon,tension=0.2,label=$r+q$,label.side=right}{v1,i2}
\fmf{dbl_curly,tension=.2,label=$q$,label.dist=15}{v1,v2}
\fmf{fermion,tension=0.2,label=$p$,label.side=right}{i3,v2}
\fmf{fermion,tension=0.2,label=$p-q$,label.side=right}{v2,i4}
\fmflabel{$g,\rho$}{i1}
\fmflabel{$g,\rho$}{i2}
\end{fmfgraph*}
\end{fmffile}
\end{center}
\begin{center}Diagram (k)\end{center}
\\
\hline
\begin{center}
\begin{fmffile}{dia6W}
\begin{fmfgraph*}(120,80)
\fmfleft{i1,i2}
\fmfright{i3,i4}
\fmflabel{$j$}{i3}
\fmflabel{$j$}{i4}
\fmf{photon,tension=0.2,label=$r$,label.side=left}{i1,v1}
\fmf{photon,tension=0.2,label=$r+q$,label.side=right}{v1,i2}
\fmf{dbl_curly,tension=.2,label=$q$,label.dist=15}{v1,v2}
\fmf{fermion,tension=0.2,label=$p$,label.side=right}{i3,v2}
\fmf{fermion,tension=0.2,label=$p-q$,label.side=right}{v2,i4}
\fmflabel{$W,Z,\rho$}{i1}
\fmflabel{$W,Z,\rho$}{i2}
\end{fmfgraph*}
\end{fmffile}
\end{center}
\begin{center}Diagram (l)\end{center}
\\
\hline
\begin{center}
\begin{fmffile}{dia6H}
\begin{fmfgraph*}(120,80)
\fmfleft{i1,i2}
\fmfright{i3,i4}
\fmflabel{$j$}{i3}
\fmflabel{$j$}{i4}
\fmf{dashes_arrow,tension=0.2,label=$r$,label.side=left}{i1,v1}
\fmf{dashes_arrow,tension=0.2,label=$r+q$,label.side=right}{v1,i2}
\fmf{dbl_curly,tension=.2,label=$q$,label.dist=15}{v1,v2}
\fmf{fermion,tension=0.2,label=$p$,label.side=right}{i3,v2}
\fmf{fermion,tension=0.2,label=$p-q$,label.side=right}{v2,i4}
\fmflabel{$_c$}{i1}
\fmflabel{$_c$}{i2}
\end{fmfgraph*}
\end{fmffile}
\end{center}
\begin{center}Diagram (m)\end{center}
\\
\hline
\end{tabular}
\end{center}
These diagrams produce the T-matrix elements,
\begin{eqnarray}
T_{f_j\,g\to f_j\,g}&=&2im_j\sum_i\frac{40 \alpha_{iSU(3)}^2}{\beta_{iSU(3)}^2}  g_{i,SU(3)}g_{ji,SU(3)} \ \frac{ E_{Gluon}}{-q^2},\\
T_{QGD,N,W}&=&2im_j\sum_i\frac{40 \alpha_{iSU(2)}^2}{\beta_{iSU(2)}^2} g_{i,W}g_{ji,W} \ \frac{ E_{W}}{-q^2},\\
T_{QGD,N,Z}&=&2im_j\sum_i\frac{40 \alpha_{iSU(2)}^2}{\beta_{iSU(2)}^2} g_{i,Z}g_{ji,Z} \ \frac{ E_{Z}}{-q^2},\\
T_{QGD,N,H}&=&2im_j\sum_i\frac{40 \alpha_{iH}^2}{\beta_{iH}^2}  g_{i,H}g_{ji,H} \ \frac{ E_{H}}{-q^2},
\end{eqnarray}
which in turn produce the differential cross sections,
\begin{equation}
\frac{d\sigma_{f_j\,g\to f_j\,g}}{d\Omega}=\frac{1}{64\pi^2 }\bigg|\sum_i\frac{40 \alpha_{iSU(3)}^2}{\beta_{iSU(3)}^2} g_{i,SU(3)}g_{ji,SU(3)}\bigg|^2 \frac{E_{Gluon}^2}{(r^2 sin^2(\theta/2))^2},
\label{sg1}
\end{equation}
and
\begin{eqnarray}
\frac{d\sigma_{QGD,N,jW}}{d\Omega}&=&\frac{(2\mu_{jW})^2}{64\pi^2 q^4}\bigg|\sum_i\frac{40 \alpha_{iSU(2)}^2}{\beta_{iSU(2)}^2}g_{i,W}g_{ji,W}\bigg|^2,\\
\label{sw1}
\frac{d\sigma_{QGD,N,jZ}}{d\Omega}&=&\frac{(2\mu_{jZ})^2}{64\pi^2 q^4}\bigg|\sum_i\frac{40 \alpha_{iSU(2)}^2}{\beta_{iSU(2)}^2}g_{i,Z}g_{ji,Z} \bigg|^2,\\
\label{sz1}
\frac{d\sigma_{QGD,N,jH}}{d\Omega}&=&\frac{(2\mu_{jH})^2}{64\pi^2 q^4}\bigg|\sum_i\frac{40 \alpha_{iH}^2}{\beta_H^2}g_{iH}g_{ji,H}\bigg|^2,
\label{sh1}
\end{eqnarray}
after using the low energy limit $E_{W,Z,H}\approx M_{W,Z,H}$.
Although these differential 
cross sections have little difference among themselves 
they represent different scattering processes: the Gluon does not carry
 mass while the 
$W,\,Z$ and Higgs do carry mass.
Therefore, (\ref{sg1}) must match the small angle deflection calculated in General Relativity which requires 
\begin{equation}
\bigg|\sum_i\frac{40 \alpha_{iSU(3)}^2}{8\pi N(1+6\alpha_{iSU(3)}^2)}\frac{g_{i,SU(3)}g_{ji,SU(3)}}{G m_j^2}\bigg| =1,
\end{equation}
and whose derivation follows exactly along the lines in section 6  above.
On the other hand the W, Z and Higgs Bosons are all massive and therefore we must compare their differential 
cross sections to those obtained from the Schroedinger equation with a standard Newtonian
potential. These 
constraints yield the following conditions
\begin{eqnarray}
\bigg|\sum_i\frac{40 \alpha_{iH}^2}{8\pi N(1+6\alpha_{iH}^2)}g_{i,H}g_{ji,H}\bigg|&=&Gm_j M_H,\\
\bigg|\sum_i\frac{40 \alpha_{iSU(2)}^2}{8\pi N(1+6\alpha_{iSU(2)}^2)}g_{i,W}g_{ji,W}\bigg|&=&Gm_j M_W,\\
\bigg|\sum_i\frac{40 \alpha_{iSU(2)}^2}{8\pi N(1+6\alpha_{iSU(2)}^2)}g_{i,Z}g_{ji,Z}\bigg|&=&Gm_j M_Z.
\label{final1}
\end{eqnarray}

\begin{center}
\begin{tabular}{ |p{7cm}| p{7cm}| }
\hline
\begin{center}
\begin{fmffile}{dia7}
\begin{fmfgraph*}(160,160)
\fmfleft{i1,i2}
\fmfright{i3,i4}
\fmflabel{$j$}{i3}
\fmflabel{$j$}{i4}
\fmf{photon,tension=0.2,label=$r$,label.side=left}{i1,v1}
\fmf{photon,tension=0.2,label=$r+q$,label.side=left}{v1,i2}
\fmf{dbl_curly,tension=.2,label=$q$,label.dist=15}{v1,v2}
\fmf{dbl_curly,label=$k$,tension=.1,label.side=right,label.dist=10}{v2,v3}
\fmf{dbl_curly,tension=.1,label=$-(k+q)$,label.side=left}{v2,v4}
\fmf{fermion,tension=0.2,label=$p$,label.side=right}{i3,v3}
\fmf{fermion,tension=0.2,label=$p-q$,label.side=right}{v4,i4}
\fmf{fermion,tension=0.2,label=$p-k$,label.side=right}{v3,v4}
\fmflabel{$m,\rho$}{i1}
\fmflabel{$m,\rho$}{i2}
\fmflabel{$^{ef}_\rho$}{v4}
\fmflabel{$^{cd}_\nu$ }{v3}
\fmflabel{$^{ab}_\mu$}{v1}
\end{fmfgraph*}
\end{fmffile}
\end{center}
\begin{center}Diagram (o)\end{center}
&
\begin{center}
\begin{fmffile}{dia8}
\begin{fmfgraph*}(160,160)
\fmfleft{i1,i2}
\fmfright{i3,i4}
\fmflabel{$j$}{i3}
\fmflabel{$j$}{i4}
\fmf{dashes_arrow,tension=0.2,label=$r$,label.side=left}{i1,v1}
\fmf{dashes_arrow,tension=0.2,label=$r+q$,label.side=left}{v1,i2}
\fmf{dbl_curly,tension=.2,label=$q$,label.dist=15}{v1,v2}
\fmf{dbl_curly,label=$k$,tension=.1,label.side=right,label.dist=10}{v2,v3}
\fmf{dbl_curly,tension=.1,label=$-(k+q)$,label.side=left}{v2,v4}
\fmf{fermion,tension=0.2,label=$p$,label.side=right}{i3,v3}
\fmf{fermion,tension=0.2,label=$p-q$,label.side=right}{v4,i4}
\fmf{fermion,tension=0.2,label=$p-k$,label.side=right}{v3,v4}
\fmflabel{$c$}{i1}
\fmflabel{$c$}{i2}
\fmflabel{$^{ef}_\rho$}{v4}
\fmflabel{$^{cd}_\nu$ }{v3}
\fmflabel{$^{ab}_\mu$}{v1}
\end{fmfgraph*}
\end{fmffile}
\end{center}
\begin{center}Diagram (p)\end{center}
\\\hline
\begin{center}
\begin{fmffile}{dia7r}
\begin{fmfgraph*}(160,160)
\fmfright{i1,i2}
\fmfleft{i3,i4}
\fmflabel{$j$}{i1}
\fmflabel{$j$}{i2}
\fmf{fermion,tension=0.2,label=$p$,label.side=right}{i1,v1}
\fmf{fermion,tension=0.2,label=$p-q$,label.side=right}{v1,i2}
\fmf{dbl_curly,tension=.2,label=$q$,label.dist=-15}{v1,v2}
\fmf{dbl_curly,label=$k$,tension=.1,label.side=left}{v2,v3}
\fmf{dbl_curly,tension=.1,label=$-(k+q)$,label.side=right}{v2,v4}
\fmf{photon,tension=0.2,label=$r$,label.side=left}{i3,v3}
\fmf{photon,tension=0.2,label=$r-q$,label.side=left}{v4,i4}
\fmf{photon,tension=0.2,label=$r-k$,label.side=left}{v3,v4}
\fmflabel{$m,\rho$}{i3}
\fmflabel{$m,\rho$}{i4}
\fmflabel{$^{ef}_\rho$}{v4}
\fmflabel{$^{cd}_\nu$ }{v3}
\fmflabel{$^{ab}_\mu$}{v1}
\end{fmfgraph*}
\end{fmffile}
\end{center}
\begin{center}Diagram (r)\end{center}
&
\begin{center}
\begin{fmffile}{dia8q}
\begin{fmfgraph*}(160,160)
\fmfright{i1,i2}
\fmfleft{i3,i4}
\fmflabel{$j$}{i1}
\fmflabel{$j$}{i2}
\fmf{fermion,tension=0.2,label=$p$,label.side=right}{i1,v1}
\fmf{fermion,tension=0.2,label=$p-q$,label.side=right}{v1,i2}
\fmf{dbl_curly,tension=.2,label=$q$,label.dist=-15}{v1,v2}
\fmf{dbl_curly,label=$k$,tension=.1,label.side=left}{v2,v3}
\fmf{dbl_curly,tension=.1,label=$-(k+q)$,label.side=right}{v2,v4}
\fmf{dashes_arrow,tension=0.2,label=$r$,label.side=left}{i3,v3}
\fmf{dashes_arrow,tension=0.2,label=$r-q$,label.side=left}{v4,i4}
\fmf{dashes_arrow,tension=0.2,label=$r-k$,label.side=left}{v3,v4}
\fmflabel{$c$}{i3}
\fmflabel{$c$}{i4}
\fmflabel{$^{ef}_\rho$}{v4}
\fmflabel{$^{cd}_\nu$ }{v3}
\fmflabel{$^{ab}_\mu$}{v1}
\end{fmfgraph*}
\end{fmffile}
\end{center}
\begin{center}Diagram (q)\end{center}
\\\hline
\end{tabular}
\end{center}

Finally,
diagrams (o)-(q) above give the desired correction to $\mathcal{O}(G^2)$ for the W, Z, and Higgs particles interacting
with Fermions:
In the low energy limit, these diagrams produce the T-matrix elements, 
\begin{eqnarray}
T_{QGD,E,B}&=&2i\frac{m_j M_B}{\sqrt{-q^2}}\bigg[\sum_i\frac{\alpha_{iB}^2}{8\beta_{iB}^2}\bigg(  51 \frac{g_{i,B}^2g_{ji,B}^2}{m_j}  +\nonumber\\
& & +\frac{\alpha_{iB}^2}{\beta_{iB}^2}\frac{g_{ji,B}g_{i,B}^3 (45M_B^2+2M_B m_j  -201 m^2_j)}{m_j^2 M_B}\bigg)\bigg],\\
\end{eqnarray}
with $B=W,\,Z,\,H$.
These T-matrices in turn produce the following differential cross sections,
\begin{eqnarray}
\frac{d\sigma}{d\Omega}_{QGD,E,B}&=&\frac{4 \mu^2_{jB}}{|q|^2} 
\bigg|\sum_i\frac{\alpha_{iB}^2}{64\pi\beta_{iB}^2} \bigg( 51 \frac{g_{i,B}^2g_{ji,B}^2}{m_j}  +\nonumber\\
& & +\frac{\alpha_{iB}^2}{\beta_{iB}^2}\frac{g_{ji,B}g_{i,B}^3 (45M_B^2+2M_B m_j  -201 m^2_j)}{m_j^2 M_B}\bigg)\bigg|^2
\end{eqnarray}
Comparison with the differential cross sections derived for a Schrodinger with a post Newtonian potential (\ref{einstein-cdbis}) imposes
\begin{eqnarray}
& &\bigg|\sum_i \frac{\alpha_{iB}^2}{64\pi\beta_{iB}^2} \bigg( 51 \frac{g_{i,B}^2g_{ji,B}^2}{m_j}  
 +\frac{\alpha_{iB}^2}{\beta_{iB}^2}\frac{g_{ji,B}g_{i,B}^3 (45M_B^2+2M_B m_j  -201 m^2_j)}{m_j^2 M_B}\bigg)\bigg|\nonumber\\
& &\;\;\;\; \;\;\;\;\;\;\;\;\;\;=2\pi a G^2 m_{j,exp}M_{B,exp}(m_{j,exp}+M_{B,exp}).
\label{final}
\end{eqnarray}
Inclusion of these differential cross sections further  increases the number of differential cross section by $12 + 2 \times (12 \times 3) = 84$ for a total of 252 
differential cross sections with a consequent increase in the total number of equations to be fitted. However, a solution can be shown 
to exist following the method in the previous section.

We could go on and construct the $\mathcal{O}(G)$ amplitude for the scattering described  by the exchange of a connecton between different Bosons
as well as the $\mathcal{O}(G^2)$ amplitude for the corresponding 1-loop corrections. 
This requires modifying $\Omega_{\mu ab}^{(i,m)}$ in (\ref{lagtodoSW}) to read 
\begin{equation}
 \Omega_{\mu ab}^{(i,m)}= \omega_{\mu ab}^{(i,m)}+ K^{(i)}_{mm'} \epsilon_{\mu ab}^{\nu}A^{m'}_\nu.
\end{equation}
The requirement that the quadratic Lagrangian (\ref{nov4}) remain invariant requires that the constant real symmetric matrices  $K^{(i)}_{mm'}$  satisfy
\begin{equation}
\sum_m K^{(i)}_{mm'} K^{(i)}_{mm''}\sim \delta_{m'm''}.
\end{equation}
The vertices (h)-(k) for Bosons emitting a connecton will then 
have a quadratic dependence on the matrices $K^{(i)}$.
Then, diagrams for Fermion-Fermion scattering remain unchanged;  diagrams for Fermion-Boson scattering will produce tree-level 
T-matrices with quadratic dependence on the matrices $K^{(i)}$;
their 1-loop corrections will produce T-matrices with quadratic and quartic dependence on the matrices $K^{(i)}$;  diagrams for Boson-Boson scattering will
produce tree-level T-matrices with quartic dependence on the matrices $K^{(i)}$; their 1-loop corrections will produce T-matrices with sextic
dependence  on the matrices $K^{(i)}$.  The coefficients of these matrices will be determined in the same manner as in the previous section: 
equating the differential
cross sections of QGD to those obtained from the Schroedinger equation and  gravitational lensing. Given the lack of experimental evidence, 
the task of writing each equation which reproduce General Relativity expectations 
is left as a curiosity for the reader.  However we note that it may very well turn out that some processes not
described by General Relativity, like  those analogous to s-channel Bhabha scattering produced by the term 
$\omega^{(i,m)}K_{mb}K_{ma}W^bG^a$, 
are confirmed through experiment or that 
some processes described by General Relativity, like those produced by the term   $\omega^{(i,m)}K_{ma}K_{ma}G^aG^a$, are confirmed to be absent through experiment.  
We stress that the matrices $K^{(i)}$ allow all these possible experimental scenarios.

\section{Black Hole Entropy and Strings}

QGD also explains the black hole entropy problem in the same way that strings do.  
To construct the map between QGD and strings we need first break down SO(4) into its isomorphic representation $SU(2)\times SU(2)$.  

We N=2 supersymmetrize at least one of QGD's SU(2) gauge groups and consider for example a K3 spacetime manifold.  
 Each SU(2) Yang Mills
 theory then has 
BPS states which have the same moduli space as that of a supersymmetric sigma model which in turn can be used to calculate the degeneracy of BPS states 
for large energy states\cite{vafa}.  The degeneracy thus obtained determines the entropy which coincides with that of certain black hole solutions in string theory
\cite{vafa}.
Thus QGD has states whose entropy is that of certain black hole solutions, and therefore, those states represent the quantum description of the black hole solutions
in the exact same way they do in String theory. 

The solution for $SU(2)$ Yang-Mills found in\cite{DV} differs from the Schwarzschild black hole solution because the singularity of its horizon 
cannot be removed through a 
coordinate transformation. However, both singularities can be located at $2GM$. Therefore in SU(2) Yang Mills, the horizon area and the
black hole mass are related in the same way as for the  Schwarzschild black hole: $A=16\pi M^2$. This result implies that the SU(2) Yang Mills black hole has
the same entropy as the Schwarzschild black hole which in turn can be expressed in terms of BPS solutions. Therefore one has a dual representation of the same
physical objects in the $N=2$ supersymmetric limit:  
the macroscopic system is described by black hole solution found in\cite{DV}, while  the microscopic system is
described by BPS states.

\section{Conclusions}

QGD, as presented here,  reproduces the expected Newtonian and post Newtonian interaction evidenced by experimental measurements 
for both matter fields and photon fields. In addition to matching the differential cross sections from QGD to those
of the Schroedinger equation with Newtonian and post Newtonian corrections as well as the differential cross section for the deflection of 
light by a point mass particle in General Relativity, QGD couples to the Standard Model in a straight forward and natural manner.  
Thus, QGD  does not require us to re-invent the Standard Model to have a unified theory.
This contrasts other attempts at describing quantum gravity which require a recasting of the 
already successful Standard Model. 

The fundamental theory, a straight forward $SO(4)$ Yang-Mills, couples to 
matter in the same manner as SU(3) and SU(2) theories describe
the other fundamental forces and thus bears a high degree of symmetry with the non-gravitational forces found in nature.
Dimensional analysis of QGD and QGD coupled to the Standard Model show that all terms in the Lagrangians have dimensionality 
four or less and therefore renormalization follows\cite{wein2}.

QGD accommodates the full particle spectrum currently
known and yields in the low energy limit the expected relativistic
forces provided a sufficiently large number of $SO(4)$ symmetries exists.  The absence in QGD or QGD coupled to the Standard Model of the graviton 
should not concern the reader; after all, we have yet to detect such a particle.

The
theory here proposed not only reproduces the observed experimental
results of General Relativity, exhibits renormalization and couples in a straight forward manner to the Standard Model,
it also incorporates yet to be observed, but highly expected, 
gravitational interactions between matter and the SU(3), SU(2) and Higgs Bosons as well as those between these Bosons.
These features make QGD coupled to the Standard Model with 
$SO(4)\times U(1)\times SU(2)\times SU(3)$ local symmetry the most attractive theory to describes all interactions in nature.

\section{Acknowledgments}
I am most grateful to J. Lee for providing {\it Ricci.m} which was indispensable for all calculations. Also indispensable was
the encouragement and funding by W. Israel during the early attempts to construct this model as well as funding by E. Braun 
in its latter stages.

\end{document}